\documentclass[12pt,a4paper]{article}
\usepackage{amssymb}
\usepackage{amsmath}
\usepackage[dvips]{graphicx}

\begin{document}
\pagenumbering{arabic}

\def\spade{\par\noindent$\spadesuit$\par\noindent}
\def\club{\par\noindent$\clubsuit$\par\noindent}
\def\heart{\par\noindent$\heartsuit$\par\noindent}
\def\G{\Gamma}

\normalsize
\rightline{IFUM-877-FT}
\rightline{November 2006}
\vskip 1 truecm
\Large
\bf
\centerline{The Hierarchy Principle and the Large Mass Limit}
\centerline{of the Linear Sigma Model}

\large
\rm
\vskip 1.3 truecm
\centerline{D.~Bettinelli\footnote{e-mail: 
{\tt daniele.bettinelli@mi.infn.it}}, 
R.~Ferrari\footnote{e-mail: {\tt ruggero.ferrari@mi.infn.it}}, 
A.~Quadri\footnote{e-mail: {\tt andrea.quadri@mi.infn.it}}}

\normalsize
\medskip
\begin{center}
Dip. di Fisica, Universit\`a degli Studi di Milano\\
and INFN, Sez. di Milano\\
via Celoria 16, I-20133 Milano, Italy
\end{center}

\vskip 0.8 truecm
\bf

\centerline{Abstract}

\normalsize
\rm

\vskip 0.5 truecm
\begin{quotation}
In perturbation theory we 
study the matching  in four dimensions between  
 the linear sigma model in the large mass limit
and the renormalized nonlinear sigma model in the recently
proposed flat connection formalism.
We consider both the chiral limit and the strong coupling limit
of the linear sigma model. 
Our formalism extends to Green functions 
with an arbitrary number of pion legs,
at one loop level,
 on the basis of the hierarchy as an efficient
 unifying principle that governs both limits. While the
chiral limit is straightforward, the matching 
in the strong coupling limit
requires careful use of the normalization conditions
 of the linear theory, in order to exploit the functional 
equation and the complete set of local solutions of its linearized form.
\end{quotation}
\newpage
\section{Introduction}
\label{sec:int}
The consistent  formulation
of the nonlinear sigma model as a finite theory by means of the
subtraction scheme based on the local functional equation
\cite{Ferrari:2005ii}-\cite{Ferrari:2005fc}
allows us to pose the question of which relation exists
between the nonlinear and the linear sigma model in the
large mass limit.
We emphasize that in our approach both theories are finite
(i.e. all divergences have been removed).

In dealing with the large mass limit we encounter
two scenarios. One is the chiral limit where
all momenta are small w.r.t. the 
vacuum expectation value (v.e.v.) $v$, the only mass scale. 
Technically we have to perform an asymptotic expansion in $1/v$ of the linear 
sigma model in the form
\begin{eqnarray}
P(\cdot,1/v)\ln(v) + Q(\cdot,1/v)
\label{int.0} 
\end{eqnarray}  
where $P,Q$ are polynomials in $1/v$. At one loop we find that the
leading terms match with the nonlinear sigma model.
The second scenario is given by the limit of strong coupling
$\lambda$.
In this latter case the momenta are not constrained 
to be small w.r.t. the scale $v$, moreover all the divergent terms in $\lambda$
have to be removed before taking the limit. The matching with the nonlinear
sigma model is again based on the leading terms, but the procedure
is more complex. In particular it shows that the use of the strong
coupling limit as a definition of the nonlinear sigma model
\cite{Bessis:1972sn},\cite{Appelquist:1980ae} leads to
a whole ensemble of intricacies related to the power law behavior in 
$\lambda$ of the v.e.v. and of the wave function renormalization constant of 
the pion fields.   

The asymptotic expansion in both scenarios is carried out
by taking the large mass limit in the Feynman integrals
by using Smirnov's technique \cite{smirnov}-\cite{smirnov2}. The two
scenarios differentiate in the use of a mass scale factor
present in the dimensional subtraction procedure.

In this paper we address the question of the large $m$ limit for the 
linear sigma model on the basis of symmetry properties associated to 
a local version of chiral transformations. This is implemented by
two important technical points. First we notice that $\tilde\Gamma$ 
(the functional one particle irreducible w.r.t. 
 the pions, but only connected w.r.t. the sigma field) 
obeys the same equation as the 1PI functional
$\Gamma_{NL}$ of the nonlinear model derived from the local 
chiral transformations. Thus $\tilde\Gamma$ is the right quantity
where to evaluate the asymptotic expansion and to impose the
matching conditions. We find it convenient to impose the same
spontaneous symmetry breaking v.e.v. and the same on-shell conditions
for the two point function of the pions in the two models. Second we use 
the hierarchical structure of the functional equation in order to study the
matching. This means that only the ancestor amplitudes need
to be studied (those with external legs given by the currents
and order parameter operator). Moreover the matching is much easier since
the number of superficially divergent ancestor amplitudes is
finite (at variance  with the amplitudes involving also the pion
field). On superficially convergent amplitudes the large mass
limit presents no difficulties, reproducing the corresponding
amplitudes of the nonlinear sigma model.

In the tree level approximation the linear sigma model
in the limit $m \rightarrow \infty$ is known to reproduce
the nonlinear theory at the first non-vanishing order
in the $1/m$ expansion \cite{Bessis:1972sn}-\cite{Sonoda:1996pd}. 
The case of the tree level approximation is automatically dealt
with by the hierarchy approach.
 
In the strong coupling limit 
particular care will be devoted in order to guarantee that the grading of 
 the perturbative expansion of the functional equation in the number of 
 loops is compatible with the asymptotic expansion in $1/m$. The removal of 
the corrections to the tadpole and to the residuum of the pion is required to 
 ensure this compatibility. This in turn
 allows to use the local invariant solutions of the linearized functional 
 equation both for $\Gamma_{NL}$ and $\tilde\Gamma$. By using the hierarchical
 principle it is then straightforward to perform the fine-tuning necessary for 
 the matching in such a way that the validity of the functional equation
is maintained through all descendant lines of the Feynman
amplitudes, in particular those involving only the Goldstone 
fields.

This strategy will be applied here at the one loop level
and, due to its generality, it could provide an effective way to study
the matching also at higher loops.

\medskip

The main results can be summarized by the following 
points. i) In the chiral limit the leading terms of  $\tilde\Gamma$ in the
linear sigma model yield the corresponding amplitudes of $\Gamma_{NL}$.
ii) In the limit of strong coupling:
ii.a) for amplitudes which are superficially convergent in the 
nonlinear sigma model the limit of large mass shows perfect matching
with the linear model; ii.b) 
for the amplitude $\tilde\Gamma_{JJ}$ with only two external background 
connections $J_{a}^{\mu}$ the matching can be  achieved 
by a fine-tuning after the subtraction
of the renormalization parts of the v.e.v. and of the pion residuum in 
the linear sigma model; ii.c) for all the other ancestor amplitudes 
$\tilde\Gamma_{K_0 K_0}$, $\tilde\Gamma_{K_0 JJ}$, $\tilde\Gamma_{JJJ}$, 
$\tilde\Gamma_{JJJJ}$ which are superficially divergent in the nonlinear 
sigma model the matching is possible provided a fine-tuning is performed 
by using the local invariant solutions of the linearized functional 
equation.\\
We stress once more that this is possible only if the compatibility
of the grading in the linearized functional equation is guaranteed
by removing  the renormalization parts
in the v.e.v. and in the residuum of the pion fields. This renormalization
removes terms in $\lambda^4 \ln\lambda$ , $\lambda^4$ and $\lambda^2$ 
which can dangerously conspire with subleading terms of the classical action
resulting in jeopardizing the hierarchy principle.\\

\medskip

The paper is organized as follows. In Sect.~\ref{sec:2} we formulate the 
linear sigma model in order to implement the local chiral symmetry in the 
 background field formalism. In Sect.~\ref{sec:3} we derive the local Ward 
 identity in the linear sigma model for the 1-PI vertex functional and for
 $\tilde\Gamma$ which is 1-PI only w.r.t. the pions fields.
  In Sect.~\ref{sec:6} we consider the tree level case 
 as a warm-up for a consistent use of the hierarchy approach
 by which the pion amplitudes are derived from those involving only
 the currents and the order parameter.
 
 In Sect.~\ref{sec:oneloop} we discuss the matching
 in the one loop approximation. 
  In Sect.~\ref{sec:4} we fix the renormalization of the linear sigma model
 and we notice the necessity to introduce a counterterm involving the field 
 strength tensor for the background connection $J_{a}^{\mu}$.
 In Sect.~\ref{sec:lm} we comment on the large
 mass expansion and discuss the different regimes
 of the chiral limit and the strong coupling limit.
 In Sect.~\ref{sec:ec} we show how to explicitly
 construct $\tilde \G^{(1)}$ and provide the general
 framework for the evaluation of the ancestor amplitudes
 at the leading order in the large mass expansion. 
  In Sect.~\ref{sec:5} we give the general proof
of the correspondence in the one loop case.   
We summarize the results for the chiral limit
in Sect.~\ref{sec:9} and for the strong coupling limit
Sect.~\ref{sec:10}. The comparison with
some previous results published in the literature
is carried out in Sect.~\ref{oth.app}.
Finally conclusions are given in Sect.~\ref{sec:11}.
 
 Appendix~\ref{app:B} gives the  Feynman rules
for the linear sigma model in the presence of a background
connection. In Appendix~\ref{app:C} we 
 give a resum\'e of the most relevant formulas.
 In Appendix~\ref{app:D} we apply the large mass
expansion due to Smirnov in order to compute
the leading and the next-to-leading term in the asymptotic expansion
of the four pion amplitude 
 and verify the cancellations of non-local terms, 
which happens at these orders as predicted by the 
hierarchy.
 In Appendix~\ref{app:G} we perform the detailed computations
 of the leading order in the large mass expansion
 for the superficially divergent ancestor amplitudes
which are needed in the general proof outlined in Sect.~\ref{sec:5}.
In Appendix~\ref{app:E} we report the classical action
of the nonlinear sigma model in the flat connection formalism
and collect the local invariant solutions of the linearized
functional equation which are needed for the large mass expansion
in the one-loop approximation. 


\section{The linear sigma model}
\label{sec:2}

The classical action of the linear sigma model can be written as
\begin{equation}\label{eq:az2}
\G^{(0)}_{L}[\phi]= \int d^Dx \, 
\Big ( \partial_{\mu}\Phi^{\dagger}\,\partial^{\mu}\Phi\,-\,
\lambda^2 v^{4-D}\,
\Big(\Phi^{\dagger}\,\Phi\,-\,\frac{v^{2}_{D}}{2}\Big)^{2} \Big ) \, ,
\end{equation} 
where $\Phi$ is the two-component complex vector field
\begin{equation}\label{eq:fiel1}
\Phi(x)=\frac{1}{\sqrt{2}}
\left( \begin{array}{c} i \phi_{1}(x) + \phi_{2}(x)\\
\phi_{0}(x) - i \phi_{3}(x)
\end{array} \right)\, . 
\end{equation}
We use a single $D$-dimensional mass scale $v_D = v^{D/2-1}$. $\lambda$ 
is dimensionless. $\G^{(0)}_L[\phi]$ in eq.(\ref{eq:az2}) is invariant
under the global chiral $SU(2)_L \times SU(2)_R$ symmetry
\begin{eqnarray}
\Phi' = U_L \Phi U_R^\dagger  ~~~~ {\rm with} ~~ U_L \in SU(2)_L \, , 
~~ U_R \in SU(2)_R \, .
\label{glob.chir}
\end{eqnarray}

The invariance under infinitesimal global $SU(2)_L$  transformations
\begin{equation}\label{eq:tra3}
\delta_{L}\Phi(x)=i\,\delta\alpha_{b}\,\frac{\tau_{b}}{2}\,\Phi(x) 
\end{equation}
is translated into the following Ward identity obeyed by $\G^{(0)}_L$:
\begin{eqnarray}
\int d^Dx \, \Big ( (\frac{1}{2} \phi_0 \delta_{ab} + \frac{1}{2} 
\epsilon_{abc} \phi_c ) 
\frac{\delta \G^{(0)}_L}{\delta \phi_b} 
- \frac{1}{2} \phi_a \frac{\delta \G^{(0)}_L}{\delta\phi_0} \Big ) =0 \, .
\label{loc.1.new}
\end{eqnarray}
Eq.(\ref{loc.1.new}) is well-known and it is usually implemented
in the literature in order to discuss the relationship
between the renormalization constants and the existence
of the Goldstone bosons.

We remark that it is possible to couple in the classical action the 
Noether current
associated with the infinitesimal left global transformations
\begin{eqnarray}\label{eq:cor4}
L^{\mu}_{a} &=& \frac{i}{2}\Big(\Phi^{\dagger}(x)\,\tau_{a}\,\partial^{\mu}
\Phi(x)\,-\,\partial^{\mu}\Phi^{\dagger}(x)\,\tau_{a}\,\Phi(x)\Big)\\\nonumber
& = & -\phi_{0}(x)\,\partial^{\mu}\phi_{a}(x)+\partial^{\mu}
\phi_{0}(x)\,\phi_{a}(x)-\epsilon_{abc}\,\partial^{\mu}\phi_{b}(x) \,
\phi_{c}(x)  
\end{eqnarray}  
to the external source $J_{\mu a}$ without violating power-counting 
renormalizability.
This yields a global Ward identity in the presence of the source $J_a^\mu$
\begin{eqnarray}
&&
\int d^Dx \, \Big ( (\frac{1}{2} \phi_0 \delta_{ab} + \frac{1}{2} 
\epsilon_{abc} \phi_c ) 
\frac{\delta \G^{(0)}_L[\phi,J]}{\delta\phi_b} 
- \frac{1}{2}  \phi_a \frac{\delta \G^{(0)}_L [\phi,J]}{\delta\phi_0} 
\nonumber \\
&& ~~~~~~~~~~~ 
- \epsilon_{abc} J^\mu_b 
\frac{\delta \G^{(0)}_L[\phi,J]}{\delta {J^\mu_c}} \Big)  = 0 
\label{loc.1.new.2}
\end{eqnarray} 
for the action
\begin{eqnarray}
\G^{(0)}_{L}[\phi,J] = \G^{(0)}_{L}[\phi] + \int d^Dx \, J_{\mu a} L^\mu_a \, .
\label{ward.1}
\end{eqnarray}
By using the fact that 
$\left . \G^{(0)}_L[\phi,J] \right |_{J=0} = \G^{(0)}_L[\phi]$
we see that for $J_{\mu a}=0$ eq.(\ref{loc.1.new.2}) 
reduces to eq.(\ref{loc.1.new}).

\section{Local Ward identity}
\label{sec:3}

The global Ward identity in eq.(\ref{loc.1.new.2}) 
does not fix the pion amplitudes in terms of the
amplitudes involving only 
the sigma field and $J^\mu_a$ 
and therefore it does not exhibit a hierarchy. 
This can be remedied by upgrading the global symmetry
to a local one
in a way compatible with power-counting renormalizability.
This can be achieved by considering the action
\begin{eqnarray}\label{eq:az2.1}
\G^{(0)} = \int d^Dx \, \Big ( D_{\mu}\Phi^{\dagger}\,D^{\mu}\Phi\,-\,
\lambda^{2} v^{4-D}\,\Big(\Phi^{\dagger}\,\Phi\,-\,\frac{v^{2}_{D}}{2}\Big)^{2}
\Big ) \, ,
\end{eqnarray}  
where $D_\mu$ denotes the covariant derivative w.r.t. $J_{\mu a}$:
\begin{eqnarray}
D_\mu \Phi = \partial_\mu \Phi - i J_{\mu a} \frac{\tau_a}{2} \Phi \, .
\label{loc.4} 
\end{eqnarray}
In this formalism $J_{\mu a}$ is a background gauge connection. In order to 
set up the perturbative expansion of the theory we expand around the
minimum constant configuration 
$\bar{\phi}_{0}\,=\,v_{D}, \bar{\phi}_{a}\,=\,0$.
Thus the field\, $\phi_{0}$\, acquires a non-vanishing v.e.v. $v_D$
and we correspondingly
shift $\phi_0$ by setting
\begin{equation}\label{eq:ext3}
\phi_{0}(x)\,=\,v_{D}\,+\,\sigma(x) \, .
\end{equation} 
The tree level Feynman rules derived from $\G^{(0)}$
are collected in
Appendix~\ref{app:B}. 

$\G^{(0)}$ is invariant under the local infinitesimal transformations
\begin{eqnarray}
&& \delta \phi_a = \frac{1}{2} (v_D + \sigma) \delta \alpha_a + 
\frac{1}{2} \epsilon_{abc} \phi_b \delta \alpha_c \, , ~~~~
\delta \sigma = -\frac{1}{2} \phi_a \delta \alpha_a \, ,
\nonumber \\
&& \delta J_{\mu a}(x) = \partial_\mu \delta \alpha_a(x) + 
\epsilon_{abc} J_{\mu b}(x) \delta \alpha_c(x) \, .
\label{loc.inf}
\end{eqnarray}
The local Ward identity fulfilled by $\G^{(0)}$ is\footnote{A subscript
denotes functional differentiation with respect to the argument.}
\begin{eqnarray}
(\frac{1}{2} (v_D + \sigma) \delta_{ab} + \frac{1}{2} \epsilon_{abc} \phi_c ) 
\G^{(0)}_{\phi_b} 
- \frac{1}{2} \phi_a \G^{(0)}_{\sigma} 
- \partial^\mu \G^{(0)}_{J^\mu_a} 
- \epsilon_{abc} J^\mu_b \G^{(0)}_{J^\mu_c} = 0 \, .
\label{loc.3.class}
\end{eqnarray}

\medskip

By adopting the normalization condition on the tadpole 
\begin{eqnarray}
\left . \G_\sigma \right |_{\sigma = \phi_a = J^\mu_a  =0} = 0 \, 
\label{loc.3.cn}
\end{eqnarray}
the position of the minimum of the potential does
not renormalize. Eq.(\ref{loc.3.class}) becomes for the full 
quantum vertex 
functional
\begin{eqnarray}
(\frac{1}{2} (v_D + \sigma) \delta_{ab} + \frac{1}{2} \epsilon_{abc} \phi_c ) 
\G_{\phi_b} 
- \frac{1}{2} \phi_a \G_{\sigma} 
- \partial^\mu \G_{J^\mu_a} 
- \epsilon_{abc} J^\mu_b \G_{J^\mu_c}  = 0 \, .
\label{loc.3.quant}
\end{eqnarray}
In our approach the matching between the linear and the nonlinear sigma model
 is studied by means of the local chiral transformations. 
In the nonlinear sigma model the $\phi_0$ field becomes a
composite operator, being subject to the nonlinear constraint
\begin{eqnarray}
\phi_0^2 + \phi_a^2 = v_D^2 \, .
\label{lm.1}
\end{eqnarray}
This suggests to introduce 
for the linear sigma model
the functional $\tilde\Gamma$ which is 1-PI
 only w.r.t. the pion fields.  For that purpose we need to perform 
the Legendre transform of the connected generating functional for the 
linear sigma model $W[K_a, K_0, J^\mu_a]$ only w.r.t. $K_a$ 
(the sources of the three independent fields $\phi_a$):
\begin{eqnarray}
\tilde \G[\phi_a, K_0, J^\mu_b]  & = &  W[K_a, K_0, J^\mu_b] - 
\int d^Dx \, K_a \phi_a \nonumber \\
& = & \G[\phi_a, \phi_0, J^\mu_b] + \int d^Dx \, K_0 \phi_0  
\label{lm.2}
\end{eqnarray}
with the Legendre transform condition
\begin{eqnarray}
\frac{\delta \G}{\delta \phi_0} = - K_0 \, 
\label{lm.2.1}
\end{eqnarray}
(i.e. $K_0$ is the source for the unshifted field $\phi_0$).

\medskip

At this stage we can derive a local functional equation for $\tilde \G$.
 From eq.(\ref{loc.3.quant}) we derive the local functional equation 
obeyed by $W[K_a,K_0,J^\mu_b]$ 
\begin{eqnarray}
&& - \frac{1}{2} \frac{\delta W}{\delta K_0} K_a 
-\frac{1}{2} \epsilon_{abc}\frac{\delta W}{\delta K_c} K_b
+ \frac{1}{2} \frac{\delta W}{\delta K_a} K_0 \nonumber \\
&& - \partial^\mu \frac{\delta W}{\delta J^\mu_a} 
   -  \epsilon_{abc} J^\mu_b \frac{\delta W}{\delta J^\mu_c} = 0 \, .
\label{f.eq.1}
\end{eqnarray}
By using the fact that
\begin{eqnarray}
\frac{\delta \tilde \G}{\delta K_0} = \frac{\delta W}{\delta K_0}
\label{f.eq.2}
\end{eqnarray}
(since we do not perform a Legendre transform w.r.t. $K_0$)
we can obtain from eq.(\ref{f.eq.1}) the following local
functional equation for $\tilde \G$%
\begin{eqnarray}
&& + \frac{1}{2} \frac{\delta \tilde \G}{\delta K_0} 
\frac{\delta \tilde \G}{\delta \phi_a}
+\frac{1}{2} \epsilon_{abc}\phi_c \frac{\delta \tilde \G}{\delta \phi_b} 
- \frac{1}{2} \phi_a K_0 \nonumber \\
&& - \partial^\mu \frac{\delta \tilde \G}{\delta J^\mu_a} 
   -  \epsilon_{abc} J^\mu_b \frac{\delta \tilde \G}{\delta J^\mu_c} = 0 \, .
\label{f.eq.3}
\end{eqnarray}
We remark that the equation for $\tilde \G$ is nonlinear,
due to the appearance of the bilinear term
in the first line of eq.(\ref{f.eq.3}) (while the
original local functional Ward identity for the linear
sigma model is linear).
Moreover by using eq.(\ref{loc.3.cn}) and eq.(\ref{lm.2}) we get
\begin{eqnarray}
\left . \frac{\delta \tilde \G}{\delta K_0} \right |_{
\phi_a = K_0 = J^\mu_a = 0} = v_D \, .
\label{f.eq.4}
\end{eqnarray}
This condition will be imposed also on the nonlinear sigma model,
thus fixing the only dimensional parameter of the theory.
We remark that eq.(\ref{f.eq.3}) is the same as the functional equation 
 obeyed by the nonlinear sigma model in the flat connection formalism 
 \cite{Ferrari:2005ii,Ferrari:2005va,Ferrari:2005fc} and
therefore we conclude that the use of  $\tilde \G$ will be the correct
way to study the matching between the two models.
In this way the matching of the limit of the linear sigma model
with the nonlinear one is performed by using the same functional
equation and the same boundary condition in eq.(\ref{f.eq.4}).

\section{Tree level results}
\label{sec:6}

We apply our strategy for the matching in $D=4$ in the tree level
approximation. In this case 
one can either compute  $\tilde \G^{(0)}$ 
by exploiting the hierarchy or directly
by means of eq.(\ref{lm.2}). 
In the first case we observe that the classical action of 
the nonlinear sigma model
$\G^{(0)}_{NL}$ (see eq.(\ref{appE:1}) in Appendix~\ref{app:E})
obeys the same functional equation (\ref{f.eq.3})
as $\tilde \G^{(0)}$.
Therefore by using the hierarchy the problem of the matching can be traced 
 back to the issue of the matching of the ancestor amplitudes
 involving the insertions of the constraint of the nonlinear sigma model 
 and of the $SU(2)$ flat connection
\begin{eqnarray}
F^\mu_a = \frac{2}{v^{2}}\big(\phi_0 \partial^\mu \phi_a - 
\partial^\mu \phi_0 \phi_a+ \epsilon_{abc} \partial^\mu \phi_b \phi_c\big) \, .
\label{tree.8}
\end{eqnarray}
In the matching procedure one has thus to properly normalize
the background connection through
\begin{eqnarray}
\label{eq.risc.1}
J_{a, NL}^\mu =-\frac{v^2}{4} J_a^\mu \, ,
\end{eqnarray}
since 
$J_{a, NL}^\mu$ couples in $\G^{(0)}_{NL}$ to
 the flat connection $F_{\mu a}$:
\begin{eqnarray}
\label{tree.7bis}
\G^{(0)}_{NL} & = & \frac{v^2}{8} \int d^4x \, 
\Big ( F^\mu_a - J^\mu_a \Big )^2 + \int d^4x \, K_0 \phi_0 \nonumber \\
              & = &  \int d^4x \, \Big ( \frac{v^2}{8} F^2 
                                         + F^\mu_a J_{\mu a,NL}
                                         + \frac{2}{v^2}  J_{NL}^2 \Big )
+ \int d^4x \, K_0 \phi_0 \, .
\end{eqnarray}
A straightforward calculation in the limit of large $m$ for $\tilde \G^{(0)}$ 
 yields in momentum space
\begin{eqnarray}
&& \tilde \G^{(0)}_{K_0}(0)=v \, ,\nonumber\\
&& \Big ( -\frac{4}{v^2} \Big )^2 
\tilde \G^{(0)}_{J^{\mu_1}_{a_1}(-p) J^{\mu_2}_{a_2}(p)} =  
\frac{4}{v^2} \delta_{a_1 a_2} g_{\mu_1\mu_2} \, . 
\label{hier.tree.1}
\end{eqnarray}
These amplitudes coincide with the corresponding ones of the nonlinear 
 sigma model
(see Appendix~\ref{app:E}), while all the remaining ancestor amplitudes 
are suppressed
 both in the chiral and in the strong coupling limit (they vanish
in the nonlinear sigma model). 
Since the ancestor amplitudes match at the first non-vanishing order, 
 the matching  at the same order for 
 amplitudes involving at least one pion field then
 follows by  the hierarchy.

In the second approach we resolve the Legendre
transform w.r.t. $\phi_0$
\begin{eqnarray}
\frac{\delta \G^{(0)}}{\delta \phi_0}  = -  K_0 \, ,
\label{tree.3} 
\end{eqnarray}
yielding
\begin{eqnarray}
- \square \phi_0 
-  J^\mu_a  \partial_\mu \phi_a - \frac{1}{2} \partial J^a \phi_a 
+ \frac{1}{4} J^2 \phi_0 
- \lambda^2 \phi_0 ( \phi_0^2 + \phi_a^2 - v^2 ) = -  K_0 \, .
\label{tree.4}
\end{eqnarray}
The above equation has to be solved for $\phi_0$ in the sense
of formal power series in $J^\mu_a, K_0, \phi_j$ and their
derivatives.
In the limit $\lambda \rightarrow \infty$ the leading contribution
comes from the non-trivial solution of
\begin{eqnarray}
 \phi_0 ( \phi_0^2 + \phi_a^2 - v^2 ) = 0 \, .
\label{tree.5}
\end{eqnarray}
The same happens in the chiral limit. Thus we find 
\begin{eqnarray}
\frac{\phi_0}{v} = \Big ( 1 - \frac{\phi_a^2}{v^2} \Big )^{1/2} + O(1/m^{2})
\, ,
\label{tree.6}
\end{eqnarray}
i.e. we recover the constraint of the nonlinear sigma model.

Then one finds that (by using the properly normalized
source $J^\mu_{a,NL}$ in eq.(\ref{eq.risc.1}))
$\tilde\G^{(0)}$ coincides in the first non-vanishing order both in 
 the chiral and in the strong coupling limit with the classical action of the 
 nonlinear sigma model $\G_{NL}^{(0)}$:
\begin{eqnarray}
&&\tilde \G^{(0)} = \G^{(0)}_{NL} + O(1/\lambda)~~~~~~~~~~~~ 
\mbox{(strong coupling limit)}\nonumber\\
&&\tilde \G^{(0)} = \G^{(0)}_{NL} + O(1/v)~~~~~~~~~~~~   
 ~~~~~~ \mbox{(chiral limit)}\, .
\label{tree.7}
\end{eqnarray}

\section{Matching by the hierarchy (one loop)}
\label{sec:oneloop}

In this Section we move to the study of the matching
in the one loop approximation.
A possible way to analyze the correspondence is
to evaluate directly $\tilde\G^{(1)}$ according to eq.(\ref{lm.2})
and verify the matching at the level of pion amplitudes
\cite{Bessis:1972sn}-\cite{Sonoda:1996pd}.

However, as we have already mentioned many times, it is more
profitable to evaluate the behavior of ancestor amplitudes
and from them derive possibly the amplitudes involving pion fields
through the hierarchy.
This strategy requires a careful analysis of the renormalization
conditions.

\subsection{Renormalization of the linear sigma model (one loop)}
\label{sec:4}

The renormalization of the linear sigma model is performed
within Dimensional Regularization. The integrals
in the Feynman amplitudes in $D$ dimensions are correctly
normalized by using the parameter $v_D$.
The local Ward identity 
(\ref{loc.3.quant}) 
restricts the one loop counterterms to be of the form
\begin{eqnarray}
\G_{ct}^{(1)} & = & \int d^Dx \, \Big [ \delta_Z (D_\mu \Phi)^
\dagger (D^\mu \Phi)
                  -\delta_t \Big(\Phi^\dagger \Phi - \frac{v^2_D}{2}\Big)
                  \nonumber \\
            & & ~~~~~~~~~~~ -\delta_\lambda \Big(\Phi^\dagger \Phi - 
\frac{v^2_D}{2}\Big)^2 + \delta_J F_{\mu\nu a} F^{\mu\nu}_{a}  \Big ] \,
\label{ren.2}
\end{eqnarray}
where $F^{\mu\nu}_a$ is the field strength of the
background connection $J^\mu_a$:
\begin{eqnarray}
F^{\mu\nu}_{a} = \partial^\mu J_a^\nu - \partial^\nu J_a^\mu + \epsilon_{abc} 
J_b^\mu J_c^\nu \, .
\label{ren.3}
\end{eqnarray}
We notice the appearance in eq.(\ref{ren.2}) of a counterterm involving the
field strength of $J_\mu^a$.
However this term does not contribute to the pion amplitudes
through the descendant lines of the hierarchy just by direct
computation (it is gauge-invariant and therefore
it disappears in the expression within
eq.(\ref{f.eq.3})).

The matching of the two models through the coincidence of the
v.e.v. requires the introduction of a counterterm for the tadpole
as in eq.(\ref{loc.3.cn}). This fixes $\delta_t$:
\begin{eqnarray}
&& \delta_t = \frac{3\lambda^2}{(4\pi)^2} m^2 \Big ( 
\frac{2}{4-D} + 1 - \gamma_E + \ln(4 \pi) - \ln \Big ( \frac{m^2}{v^2} \Big )
\Big ) \, .
\label{ren.t}
\end{eqnarray}
By the above condition we are able to fix a common mass scale 
 for both the linear and the nonlinear sigma model 
 according to eq.(\ref{f.eq.4}) which at one loop gives\\
\begin{eqnarray}
\tilde \G^{(1)}_{K_0}=0\, . 
\label{ren.tilde1}
\end{eqnarray}
We also require that the residuum of the pion is 
set equal to one, thus fixing $\delta_Z$:
\begin{eqnarray}
\delta_Z = - \frac{\lambda^2}{(4\pi)^2} \, . 
\label{ren.z}
\end{eqnarray}
We remark that if one does not remove the power law dependence 
in $\lambda$ and $v$ of 
 $\G^{(1)}_\sigma$ and  $\G^{(1)}_{JJ}$ by using the counterterms in 
 eqs.(\ref{ren.t}) and (\ref{ren.z}) the suppressed terms 
in $1/\lambda$ or $1/v$ (according to which limit is taken) 
entering in $\tilde\G^{(0)}$ (see eq.(\ref{tree.7})) would spoil 
through the bilinear terms in eq.(\ref{f.eq.3}) the compatibility between 
the large mass limit and the loop expansion.
A similar problem is also encountered in the direct
evaluation of pion amplitudes by conventional
approaches discussed in Sect.~\ref{oth.app}.

In our strategy of using only ancestor amplitudes 
the condition on the residuum of the pion can be translated in terms
of $\tilde \G^{(1)}_{JJ}$ as follows:
\begin{eqnarray}
\lim_{m^2 \to \infty} \frac{\partial}{\partial m^2} 
\tilde \G^{(1)}_{JJ} = 0 \, .
\label{ren.tilde}
\end{eqnarray}

The remaining counterterms are chosen according to Minimal
Subtraction on the basis of simplicity and elegance:
\begin{eqnarray}
&& \delta_\lambda = \frac{12 \lambda^4}{(4 \pi)^2} \frac{2}{4-D} \, , 
\nonumber \\
&& \delta_J = \frac{1}{24 (4 \pi)^2} \frac{2}{4-D} \, .
\label{ren.4}
\end{eqnarray}

\subsection{Large mass expansion}\label{sec:lm}

The expansion for large value of the parameters is performed
by using the technique devised by Smirnov \cite{smirnov}-\cite{smirnov2}
and it involves in principle only one parameter, the mass
of the heavy particle running inside the graph.
However, if the graph is divergent at $D=4$, then the pole
subtraction introduces a second mass which restores
the correct dimensions of the Feynman amplitudes.
If this extra mass scale is kept fixed for $m\rightarrow \infty$
we realize the strong coupling limit; if instead  this mass scale
is identified with the v.e.v. we obtain the chiral limit. In both cases
the technical job is the same and consists in asymptotic expansion
of the Feynman amplitudes for large $m$.

In Appendix~\ref{app:G} we evaluate the asymptotic expansion
of all the ancestor amplitudes of the linear sigma model
which develop a singular behavior in the limit $m \rightarrow \infty$
and which correspond to superficially divergent amplitudes of the
nonlinear sigma model.

>From the results of Appendix~\ref{app:G} one sees that
in the chiral limit the correspondence with the nonlinear
sigma model is automatic if one keeps only
the leading terms in $\ln v$.
In particular all graphs involving a virtual sigma field
yield subleading contributions.

In the strong coupling limit the connected graphs
containing the sigma field ($\Delta \tilde \G^{(1)}$) yield
terms proportional to $\ln \lambda$ which have to be subtracted
before taking the limit $\lambda \rightarrow \infty$.

In performing this kind of calculations one finds a certain
number of cancellations which can be traced from the
algebraic characterization of $\tilde \G^{(1)}$.
In the next subsection we are going to elaborate
on this more general and formal approach which has
the advantage of being able to deal with
the general case with any number of legs in the
ancestor amplitudes. On the account of the hierarchy that means that 
by this method
we can take the limit of large mass for any
Feynman amplitude with arbitrary number of pion fields (one loop).

\subsection{Explicit construction of $\tilde \G^{(1)}$}\label{sec:ec}

At one loop it is possible to give an analytic construction of
$\tilde \G^{(1)}$ by using $\G^{(1)}$ and by making use
of the Legendre transform in eq.(\ref{lm.2}).
For that purpose we expand
the solution of
eq.(\ref{lm.2.1}) as
\begin{eqnarray}
\phi_0 = \phi_0^{(0)} + \phi_0^{(1)}  + \dots \, ,
\label{match.1}
\end{eqnarray}
where $\phi_0^{(1)}$ stands for the one loop corrections to the solution
of the classical equation of motion in eq.(\ref{tree.4})
and the dots
denote terms of higher order in $\hbar$.
By substituting $\phi_0$ in eq.(\ref{match.1})
into eq.(\ref{lm.2}) and by keeping only terms up to order
one in the loop expansion one finds 
\begin{eqnarray}
&& \tilde \G[\phi_a,K_0,J_{a\mu}]  = 
\G^{(0)}[\phi_a,\phi^{(0)}_0,J]
+  \int d^Dx \, K_0 \phi_0^{(0)} \nonumber \\
&&~~~~~~ + \G^{(1)}[\phi_a,\phi^{(0)}_0,J]+ \int d^Dx\,  
\Big ( \frac{\delta \G^{(0)}[\phi_a,\phi^{(0)}_0,J] }{\delta \phi_0} 
+ K_0 \Big ) \phi_0^{(1)} + \dots \nonumber \\
&&~~~~~~ =  \G^{(0)}[\phi_a,\phi^{(0)}_0,J]
+  \int d^Dx \, K_0 \phi_0^{(0)} +  \G^{(1)}[\phi_a,\phi^{(0)}_0,J] + \dots
\label{match.2}
\end{eqnarray}
where use has been made of eq.(\ref{tree.3}). 
By eq.(\ref{match.2}) $\tilde \G^{(1)}$ can be obtained by
 substituting in the one loop 1-PI vertex functional $\G^{(1)}$ of the linear 
 sigma model the solution of the classical Legendre transform in 
 eq.(\ref{tree.3}):
\begin{eqnarray}
\tilde \G^{(1)}[\phi_a, K_0, J_{a \mu}] = \left .
\G^{(1)}[\phi_a, \phi_0, J_{a\mu}] 
\right |_{\phi_0 = \phi_0^{(0)}(\phi_a, K_0, J_{a\mu})} \, .
\label{tildeg1}
\end{eqnarray}

$\tilde\G^{(1)}$ obeys the linearized functional equation obtained by 
 projecting eq.(\ref{f.eq.3}) at one loop
\begin{eqnarray}
&& + \frac{1}{2} \frac{\delta \tilde \G^{(0)}}{\delta K_0} 
\frac{\delta \tilde \G^{(1)}}{\delta \phi_a} + \frac{1}{2} 
\frac{\delta \tilde \G^{(1)}}{\delta K_0} 
\frac{\delta \tilde \G^{(0)}}{\delta \phi_a}
+\frac{1}{2} \epsilon_{abc}\phi_c \frac{\delta \tilde \G^{(1)}}{\delta \phi_b} 
 \nonumber \\
&& - \partial^\mu \frac{\delta \tilde \G^{(1)}}{\delta J^\mu_a} 
   -  \epsilon_{abc} J^\mu_b \frac{\delta \tilde \G^{(1)}}{\delta J^\mu_c} 
= 0 \, .
\label{l.eq.3}
\end{eqnarray}

Eq.(\ref{l.eq.3}) suggests a different approach to the
study of the matching than the direct evaluation
of the pion amplitudes. Since eq.(\ref{l.eq.3}) implements
a hierarchy principle, the coincidence in the large
mass limit of the ancestor amplitudes is sufficient to
guarantee through the hierarchy 
 the coincidence (at the same order in the large mass expansion)
of amplitudes involving at least one pion.  

According to eq.(\ref{tildeg1})
in order to derive the 
one loop ancestor amplitudes at the first
non-vanishing order in the large mass expansion
one has to evaluate the renormalized 1-PI vertex functional
$\G^{(1)}[0,\sigma,J]$ of the linear sigma model at $\phi_a=0$
(no external pion legs) and then substitute
into it the solution of the Legendre transform in eq.(\ref{tree.4}),
again by keeping only those terms non-vanishing at $\phi_a=0$.
In the first order of the large $m$ expansion the solution
to eq.(\ref{tree.4}) can be truncated
at order $1/\lambda^2$ (further terms yield
suppressed contributions in the large $m$ expansion):
\begin{eqnarray}
\sigma(x) = \frac{1}{8 \lambda^2 v} J^2(x) + \frac{1}{2\lambda^2 v^2} K_0(x)
+ \dots
\label{1l.2}
\end{eqnarray}

\subsection{General proof of the matching}
\label{sec:5}

In this subsection we exploit eq.(\ref{tildeg1})
in order to derive a general framework
for the analysis  of the ancestor
amplitudes in $\tilde \G^{(1)}$.

For that purpose we consider first the large mass limit
of $\G^{(1)}[0,\sigma,J]$.
As can be seen from the tree-level Feynman rules
in Appendix~\ref{app:B}, the degree in $\lambda$ and $v$
of the interaction vertices only depends on the 
number of fields entering into the vertex (irrespective
of whether the fields are $\sigma$ or $\phi_a$).

As a consequence, the leading order in the
large mass expansion for a superficially convergent
1-PI amplitude in $\G^{(1)}[0,\sigma,J]$ is given by graphs
with only massless internal lines (light
graphs). In fact  any graph with superficial degree of divergence
$\delta <0$ contributing to such an amplitude and involving  
at least one massive virtual sigma field  
(heavy graph) 
is suppressed  by a factor 
$m^\delta$ w.r.t. the light graphs.

For superficially divergent amplitudes the same conclusion
holds in the chiral limit, as can be checked by the explicit
computations of the relevant amplitudes in Appendix~\ref{app:G}
once the normalization of the vector external source in
eq.(\ref{eq.risc.1}) is taken into account.

On the contrary, in the strong coupling limit the dominant
contribution to superficially divergent amplitudes
comes from the heavy graphs.
The leading terms for large $\lambda$ 
can be computed by using the results of 
Appendix~\ref{app:G} and yield
(the subscript $H$ stands for heavy):
\begin{eqnarray}
&& \!\!\!\!\!\!\!\!\!\!\!\!\!\!\!
\G^{(1)}_{H}  =  
\frac{1}{(4\pi)^2} \ln \Big ( \frac{m^2}{v^2} \Big )
 \int d^4x \,
\Big ( -9 \lambda^4 v^2 \, \sigma^2 
- \frac{1}{4} \lambda^2 v \, J_a^2 \sigma 
+ \frac{1}{24}(\partial_\mu J_{a\nu} \partial^\mu J_a^\nu  -
\partial J_a \partial J_a ) \nonumber \\
&& ~~~~~~~~ - \frac{1}{8} \epsilon_{abc}
\partial_\mu J_{a\nu} J^\mu_b J^\nu_c 
- \frac{1}{96} ( J_a^2 J_b^2 + 2 J_{a\mu} J_b^\mu J_{a\nu} J_b^\nu )
+\frac{3}{64}  J_a^2 J_b^2 \Big ) \, . 
\label{appBB:1} 
\end{eqnarray}
On account of the replacement in eq.(\ref{1l.2})
$\G^{(1)}_H$ gives rise through the Legendre transform 
in eq.(\ref{tildeg1}) to logarithmically divergent 
terms in the ancestor amplitudes $\tilde \G^{(1)}_{K_0 K_0}$,
$\tilde \G^{(1)}_{K_0 JJ}$, $\tilde \G^{(1)}_{JJ}$,
$\tilde \G^{(1)}_{JJJ}$ and $\tilde \G^{(1)}_{JJJJ}$.
These amplitudes are in one-to-one correspondence with the one loop
superficially divergent ancestor amplitudes
of the nonlinear theory.
The logarithmic divergences in $\lambda$
have to be removed in a symmetric way
by making use of the invariants ${\cal I}_1, \dots, {\cal I}_7$
in eq.(\ref{appE:4}) before the limit $\lambda \rightarrow \infty$
is taken.

Once logarithmic divergences are removed, 
the finite parts generated by the heavy graphs in
the large $\lambda$ expansion have finally to be
matched by 
a finite  fine-tuning by using  the invariants ${\cal I}_1, \dots, {\cal I}_7$.
In fact at one loop level the linearized
equation (\ref{l.eq.3}) guarantees that this is always
a viable procedure compatible with allowed one-loop
finite renormalizations of the nonlinear sigma model.

The recursive fulfillment of the bilinear 
local functional equation (\ref{f.eq.3})
order by order in the loop expansion
is required in order to guarantee the locality
of the symmetric counterterms  \cite{Picariello:2000xc}.
At order $n>1$ the latter obey the
inhomogeneous equation  obtained
by projecting the bilinear functional
equation at order $n$ in the loop expansion.
The higher order symmetric 
counterterms can be determined by using the 
algebraic methods 
based on the Slavnov-Taylor (ST)  parameterization
of the symmetric effective action which were
originally developed for the restoration
of the ST identities 
in chiral gauge theories in the absence of a symmetric regularization 
\cite{Quadri:2005pv}- \cite{Quadri:2003ui}.

The fact that the subtraction of the divergent terms
cannot be defined in a unique way shows that
the nonlinear sigma model cannot be uniquely defined
as a limit of strong coupling.

\medskip

In order to establish the matching with the nonlinear sigma model
it remains to prove that the light graphs in $\G^{(1)}[0,\sigma,J]$
do reproduce under the Legendre transform
in eq.(\ref{tildeg1}) the same ancestor amplitudes of the nonlinear
theory (at the leading order). 
This result relies on quite subtle cancellations implemented 
in $\tilde \G^{(1)}$ by the Legendre transform.
In particular the contributions generated by the quadrilinear
vertex $\G^{(0)}_{JJ\phi\phi}$ (which is absent in the
nonlinear sigma model) are removed only once the Legendre
transform is performed, as shown explicitly
later on. Some examples of
these cancellations have been pointed out
in Appendix~\ref{app:G} (see eqs.(\ref{eq:four10}),
(\ref{eq:four14}) and (\ref{eq:four18})).

Here we elucidate the general structure of the 
underlying cancellation mechanism.

The evaluation  of the ancestor amplitudes 
in $\tilde \G^{(1)}$
at the required order
is carried out by properly  normalizing the relevant light amplitudes
in $\G^{(1)}[0,\sigma,J]$  with the help of the single mass scale $v$ 
and then by performing the replacement in eq.(\ref{1l.2}) on the subtracted
amplitudes. Since in this process
$\sigma$ plays the r\^ole of an external background source,
it is convenient to derive an effective potential
in such a way that the effects of the replacement 
are taken into account directly at the level of the Feynman rules. 
One is then led to introduce the potential $V$ given by
\begin{eqnarray}
\!\!\!\!\!\!\!\!\!
V(\phi_a;K_0 , J) & = & \left .
\int d^Dx \, \Big ( \frac{1}{2} \G^{(0)}_{\sigma \phi_{a_1} \phi_{a_2}}
\sigma \phi_{a_1} \phi_{a_2} 
+ \frac{1}{2} \G^{(0)}_{J^\mu_a \phi_{a_1} \phi_{a_2}} 
J^\mu_a  \phi_{a_1} \phi_{a_2} \right . \nonumber \\
& & ~~~~~ \left .  
+ \frac{1}{4} \G^{(0)}_{J^{\mu_1}_{a_1} 
J^{\mu_2}_{a_2} \phi_{b_1} \phi_{b_2}} J^{\mu_1}_{a_1} 
J^{\mu_2}_{a_2} \phi_{b_1} \phi_{b_2}
\Big ) \right |_{ \sigma = \frac{1}{8 \lambda^2 v^{3-D/2}} 
J^2 + \frac{1}{2\lambda^2 v^2} K_0} \, . \nonumber \\
\label{appBB:2.bis}
\end{eqnarray}
Several comments are in order here.
The interaction vertices between round brackets in 
eq.(\ref{appBB:2.bis})
 are precisely those generating the light graphs
 in  $\G^{(1)}[0,\sigma,J]$
  (through  contraction with the pion propagators).
 Moreover  it is important to notice that,
 as a consequence of the use of a single mass scale $v$, 
the replacement for $\sigma$ 
 in eq.(\ref{appBB:2.bis}) can 
be safely carried out in $D$ dimensions.

Thus we can formally write
\begin{eqnarray} 
&& 
\!\!\!\!\!\!\!\!\!\!\!\!\!\!\!\!\!\!
\exp \Big ( \frac{i}{\hbar} 
\tilde \G^{(1)}[K_0,J] \Big ) \simeq_\hbar
\exp \Big ( \frac{i}{\hbar} \int d^Dx \, V 
\Big ( \frac{\hbar}{i} \frac{\delta}{\delta K_a}; 
K_0, J_{\mu a} \Big ) \Big )  \nonumber \\
&&  \left . 
\exp \Big ( - \frac{1}{2\hbar} \int d^Dx \, K_a \Delta_{ab} K_b \Big )  
\right |_{K_a = 0} \, ,
\label{appBB:2}
\end{eqnarray}
where $\Delta_{ab}$ is the pion propagator
\begin{eqnarray}
\Delta_{ab} = \frac{i}{p^2} \delta_{ab} \, .
\label{pion.prop}
\end{eqnarray}

The equality in eq.(\ref{appBB:2})
holds at the leading order in the large mass
expansion. The subscript $\hbar$
states that one has to keep only terms of order $\hbar$ in the R.H.S.
of eq.(\ref{appBB:2}). 
The amplitudes generated according to eq.(\ref{appBB:2})
have to be properly normalized by using the single mass scale
$v$ before subtracting the pole. Notice that 
since each interaction vertex in $V$  contains
two fields $\phi_a$, the one-loop connected graphs with
external legs $K_0$ and $J$ are automatically 1-PI.

By substituting $\sigma$ as displayed in
eq.(\ref{appBB:2.bis}) the effective potential $V$ reads
explicitly
\begin{eqnarray}
&& \!\!\!\!\!\!\!\!\!\!\!\!\!\!\!\!
 V(\phi_a; K_0, J)
 =  \nonumber \\
& &  
\!\!\!\!
\int d^Dx \, \Big ( \frac{1}{4 \lambda^2 v^2 } 
\G^{(0)}_{\sigma \phi_{a_1} \phi_{a_2}} K_0
 \phi_{a_1} \phi_{a_2} 
+ \frac{1}{2} \G^{(0)}_{J^\mu_a \phi_{a_1} \phi_{a_2}} 
J^\mu_a  \phi_{a_1} \phi_{a_2} \nonumber \\
& &  
\!\!\!\!
+\Big ( \frac{1}{4}   \G^{(0)}_{J^{\mu_1}_{a_1} 
J^{\mu_2}_{a_2} \phi_{b_1} \phi_{b_2}} 
+ \frac{1}{16 \lambda^2 v^{3-D/2}} \G^{(0)}_{\sigma \phi_{b_1} \phi_{b_2}} 
g_{\mu_1 \mu_2} \delta_{a_1 a_2} \Big )
J^{\mu_1}_{a_1} J^{\mu_2}_{a_2} \phi_{b_1} \phi_{b_2}\Big )  \, .
\label{appBB:4}
\end{eqnarray} 
Since
\begin{eqnarray}
\G^{(0)}_{\sigma \phi_{a_1} \phi_{a_2}} = - 2 \lambda^2 v^{3-D/2} 
\delta_{a_1 a_2}
\label{appBB:5}
\end{eqnarray} 
one gets 
\begin{eqnarray}
\int d^Dx \, \frac{1}{4 \lambda^2 v^2 } 
\G^{(0)}_{\sigma \phi_{a_1} \phi_{a_2}} K_0
 \phi_{a_1} \phi_{a_2} = -\int d^Dx \, \frac{1}{2v_D} K_0 \phi_a^2 \, 
 \label{appBB:6}
\end{eqnarray}
i.e. the same trilinear coupling $K_0\phi\phi$
appearing in $\G^{(0)}_{NL}$.
The trilinear coupling $J\phi\phi$ in eq.(\ref{appBB:4}) coincides
with the corresponding coupling in $\G^{(0)}_{NL}$
(once the source $J_{a\mu}$ is properly rescaled
as in eq.(\ref{eq.risc.1})).
Moreover 
\begin{eqnarray}
 \frac{1}{4}   \G^{(0)}_{J^{\mu_1}_{a_1} J^{\mu_2}_{a_2} \phi_{b_1} 
\phi_{b_2}} + \frac{1}{16 \lambda^2 v^{3-D/2}} 
\G^{(0)}_{\sigma \phi_{b_1} \phi_{b_2}} 
g_{\mu_1 \mu_2} \delta_{a_1 a_2} = 0 \, 
\label{appBB:7}
\end{eqnarray}
and therefore the quadrilinear coupling $JJ\phi\phi$ in eq.(\ref{appBB:4})
 vanishes. 

Thus we see that the effective potential $V$ coincides
with the one of the nonlinear sigma model.
Since we use a single mass scale $v$ for the normalization
of both $\tilde \G^{(1)}$ and $\G^{(1)}_{NL}$, we can then state 
the coincidence of the one-loop convergent ancestor amplitudes 
(at the leading order) in both theories.
For the divergent ones, as discussed before, finite parts
need to be matched by a fine-tuning by using
the invariants ${\cal I}_1, \dots, {\cal I}_7$ since
the subtraction of the divergent parts is not uniquely
defined.

This property implies that also the amplitudes involving 
at least one pion field,
generated through the descendant lines of the hierarchy, 
 coincide at the leading order,
thus establishing the matching in full generality (one loop). 

\section{One loop chiral limit}
\label{sec:9}

In the chiral limit at the leading order (logarithmic dependence
on $v$ and powers in $1/v$) the ancestor amplitudes of $\tilde \G^{(1)}$ 
coincide with those of the nonlinear theory.
In fact after fixing the v.e.v. of the order parameter by
eq.(\ref{ren.t}) and by using
the normalization of the pole of the $\phi_a$ fields imposed in 
eq.(\ref{ren.z}) 
no further fine tuning is necessary in order to obtain the 
matching. 
The correspondence with the nonlinear sigma model amplitudes  
is guaranteed provided that one uses a single subtraction mass scale $v$.

According to the discussion of the previous Section, the matching
of the ancestor amplitudes is realized in a substantially 
different way depending on whether the amplitude
is superficially divergent or convergent.
If the amplitude is convergent then the limit can be performed
in a straightforward way by expanding in $1/v$
and the coincidence of the two models at the leading order
is automatically fulfilled.

For divergent ancestor amplitudes the discussion is more involved.
First of all we notice that on the side of the linear sigma model
the behavior in $\ln \frac{m^2}{v^2}$ of the large mass
expansion of the heavy graphs $\Delta \tilde \G^{(1)}$ 
(those of $\tilde \G^{(1)}$ which have no corresponding
partner amplitudes in $\G^{(1)}_{NL}$) is a constant
since $m^2 = 2 \lambda^2 v^2$, as shown in Appendix~\ref{app:G}.
Thus at the one loop level the contributions of the heavy
superficially divergent amplitudes are of order zero w.r.t.
superficially divergent amplitudes of the nonlinear sigma model,
which behave as $\ln v$.

However these finite terms are not relevant for the matching.
The reason is due to the fact that $\tilde \G^{(1)}$ satisfies
the functional equation in the linearized form of eq.(\ref{l.eq.3})
and therefore any finite linear combination of local invariants
${\cal I}_1, \dots, {\cal I}_7$ in eq.(\ref{appE:5})
can be added to $\G^{(1)}_{NL}$ as an allowed finite renormalization.
This voids of any relevance the terms
$\Delta \tilde \G^{(1)}$.

Once more we point out that this kind of argument can be used
only for the ancestor amplitudes. 
In fact there are infinitely many 
divergent descendant amplitudes. 
The latter are however related through the hierarchy
to a finite number of ancestor amplitudes only.

Our results on the chiral limit of the linear sigma model reproduce those 
 obtained in chiral perturbation theory in 
Refs.\cite{CHPT,CHPT1} once one uses the (one-loop) 
correspondence table between chiral invariants and the invariants 
${\cal I}_1, \dots , {\cal I}_7$ of the 
 linearized functional equation given in \cite{Ferrari:2005va}.

However this agreement is presumably limited to the
one loop approximation. In our opinion the use of the
tree level equations of motions in Refs.\cite{CHPT,CHPT1}
instead of the Legendre transform from the vertex
functional to the connected generating functional
is only valid at one loop.
The extension of the latter method to higher orders is less 
straightforward since
higher orders (non-local) corrections to the tree-level 
equations of motion have to be
taken into account. 
In the one-loop approximation the use of the tree-level 
equations of motion is very advantageous since it allows
to perform the matching directly on the connected Green
functions of the linear theory, as in done in
Refs.\cite{CHPT,CHPT1}.

\section{One loop  $\lambda \to \infty$ limit}
\label{sec:10}

The strong coupling limit turns out to be more involved. 
The limit $\lambda \to \infty$ is singular as it is evident 
from the logarithmic dependence on $\lambda$ of the
functional $\G^{(1)}_H$ in eq.(\ref{appBB:1}).
This residual logarithmic dependence cannot be  removed
by a renormalization of the linear sigma model
(since the required counterterms violate the power-counting bounds) and
therefore have to be introduced as counterterms
for the effective action $\tilde \G^{(1)}$ by means
of the invariants ${\cal I}_1, \dots, {\cal I}_7$
in eq.(\ref{appE:4}). This is in contrast with the divergent
term proportional to $\lambda^2$ in the two-point
function $\tilde \G^{(1)}_{JJ}$, which can be removed
by using the tadpole counterterm in eq.(\ref{ren.t}) and 
the  kinetic counterterm in eq.(\ref{ren.z}) introduced in order to fix the 
residuum of the $\phi_a$ propagator to one.
The removal
procedure has to be consistent
with the one loop local functional equation,
therefore we have expanded the $\ln \lambda$ terms as linear
combinations of the complete basis given by ${\cal I}_1, \dots, {\cal I}_7$.
By comparison with eq.(\ref{appBB:1}) 
and by taking into account eq.(\ref{appE:4})
one finds\\
\begin{eqnarray}\label{eq:coun}
\tilde{\G}^{(1)}  & = & 
-\frac{1}{4(4\pi)^2}\ln\Big(\frac{m^2}{v^2}\Big)
\Big[-\frac{1}{6}({\cal I}_1 -{\cal I}_2) +\frac{1}{2}{\cal I}_3
\nonumber \\
& & ~~~~~~~~~
+\frac{9}{v^4} {\cal I}_4 
+\frac{5}{v^2}{\cal I}_5 + \frac{1}{24}({\cal I}_6 +2{\cal I}_7)
+\frac{1}{2} {\cal I}_6\Big] +O(\lambda^0)\, .
\end{eqnarray}
Once the logarithmic divergences are removed, the finite parts
generated by the heavy graphs in the strong coupling limit
have to be matched by using the invariants ${\cal I}_1, \dots, {\cal I}_7$.
The coincidence of the ancestor amplitudes then guarantees
through the hierarchy the matching of the pion amplitudes.

As we have already said, the
lack of uniqueness in the subtraction procedure 
reflects the impossibility of uniquely
defining the nonlinear sigma model as the strong coupling limit
of the linear model.

\medskip

As an application of the hierarchy principle, we determine the behavior 
 of the four point pion function
in the strong coupling limit. 
For that purpose one has to project
eq.(\ref{eq:coun}) on the monomials containing four pion fields.
By using eq.(\ref{appE:4}) one finds
\begin{eqnarray}
\!\!\!\!\!\!\!\!\!\!\!\!\!\!\!\!
{\tilde \G}^{(1)}[\phi\phi\phi\phi] & = &
-\frac{1}{2(4\pi)^2} \frac{\ln\lambda}{v^4}
\Big [ \frac{8}{3} (\partial_\mu \phi_a \partial^\mu \phi_a 
\partial_\nu \phi_b \partial^\nu \phi_b - \partial_\mu \phi_a 
\partial_\nu \phi_a \partial^\mu \phi_b \partial^\nu \phi_b) \nonumber \\
&&  + 9 \phi_a \square \phi_a \phi_b \square \phi_b
+ 20 \phi_a \square \phi_a \partial_\mu \phi_b \partial^\mu \phi_b
\nonumber \\
&&  + \frac{26}{3} \partial_\mu \phi_a \partial^\mu \phi_a 
\partial_\nu \phi_b \partial^\nu \phi_b 
+ \frac{4}{3}   \partial_\mu \phi_a \partial^\mu \phi_b 
\partial_\nu \phi_a \partial^\nu \phi_b \Big ] + O(\lambda^0)\, . 
\label{eq.param.1}
\end{eqnarray}

\subsection{Comparison with some previous results present in the literature}
\label{oth.app}

\begin{figure}[h]
\begin{center}
\includegraphics[width=0.6\textwidth]{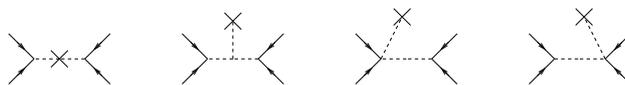}
\end{center}
\caption{Counterterm contributions to the four point function}
\label{fig.count}
\end{figure}

The comparison of the above result with previous published works, for 
instance Ref.~\cite{Appelquist:1980ae}, is complicated by the fact  
that the conditions in eqs. (\ref{ren.tilde1}) and (\ref{ren.tilde}) 
are not imposed on the linear sigma model. The counterterms needed 
to restore these conditions ($\delta_t$ and $\delta_Z$) have a $\lambda^2$,
 $\lambda^4$ and $\lambda^4 \ln\lambda$ behavior. The power factors
in $\lambda$ might cause important changes in the four-pion
amplitudes. We comment here on the relevance of these changes.\\ 
Apparently the subtraction procedure used by Appelquist and Bernard
\cite{Appelquist:1980ae} differs
from ours as can be seen from their radiative correction to the v.e.v.\\
\begin{eqnarray}
 - \frac{3}{32 \pi^2} \frac{m^4} {v} \big(\ln(m^2) - 1 + \gamma_E + 
\ln\pi \big) \, .
\label{AB}
\end{eqnarray}
A consistent use of this counterterm together with the one in 
eq.(\ref{ren.z}) in the relevant graphs depicted in Figure~\ref{fig.count} 
shows that in the four-pion
amplitudes  the $\lambda^4$ terms disappear, however the $\lambda^2$ and  
$\lambda^2 \ln\lambda$ are non zero. These cancel exactly against the 
corresponding terms reported in Ref.~\cite{Appelquist:1980ae} and only 
$\ln\lambda$ terms are left over. The final result matches with 
eq.(\ref{eq.param.1}), thus showing that it is crucial to keep the
v.e.v. of sigma and the residuum of the pion pole fixed in the renormalization
of the linear sigma model, if one takes the limit of strong coupling.

\section{Conclusions}
\label{sec:11}

In this paper we have examined the matching between 
the linear sigma model and the nonlinear sigma model
in $D=4$ by using the symmetry properties based on the
local chiral transformations.
By formulating the linear sigma model in terms of a
background connection we have derived
a linear local functional equation for the linear
sigma model vertex functional. 
In order to study the matching in the large $m$
limit one has to consider amplitudes which are 
1-PI w.r.t. the pion fields but connected w.r.t. the
sigma lines. These amplitudes are collected in the generating functional
$\tilde \G$. We have shown that a nonlinear local functional
equation for $\tilde \G$ holds. This equation is of the
same functional form as the one fulfilled by  the
nonlinear sigma model vertex functional in the flat
connection formalism.

By fixing a common mass scale 
for the linear and the nonlinear theory
(the v.e.v. of the order parameter)
and by adopting the normalization condition
on the residuum of the pions (which ensures
 the compatibility between the large mass limit and
the loop expansion at the level of the functional equation) 
we have shown that the same hierarchical structure 
exists both in the linear and the nonlinear sigma model.

This allows us to derive the matching for any amplitude involving at
least one pion once the matching is verified at the level
of ancestor amplitudes only (i.e. amplitudes involving
the insertions of the flat connection and of the constraint
of the nonlinear sigma model).

This provides a common framework for the study of both the
chiral limit and the strong coupling limit.
 
In the first case we have shown in the one loop approximation that the 
matching is fulfilled at the leading order (logarithmic dependence on $v$
 and powers in $1/v$)
provided that the normalization conditions on the tadpole
(fixing the mass scale) and on the residuum of the pion
are imposed on the linear side.

In the strong coupling limit we find that the limit
$\lambda \rightarrow \infty$ is logarithmically
divergent at one loop and thus the removal of these
divergences have to be performed before
the limit is taken.
The logarithmic divergences have to be subtracted symmetrically
through the invariant solutions of the linearized
functional equation.

Once the logarithmic divergences are removed, the finite parts
generated by the heavy graphs in the strong coupling limit
have to be matched by using the invariants ${\cal I}_1, \dots, {\cal I}_7$
in eq.(\ref{appE:4}).
The coincidence of the ancestor amplitudes then guarantees
through the hierarchy the matching of the descendant amplitudes
involving at least one pion.

The lack of uniqueness of the subtraction procedure 
reflects the impossibility of uniquely
defining the nonlinear sigma model as the strong coupling limit
of the linear model.

We remark that 
in the case of the strong coupling limit the hierarchy provides an 
efficient way to control the logarithmic dependence 
of all one-loop amplitudes 
on the coupling constant
through the functional $\G^{(1)}_H$ given in eq.(\ref{appBB:1}).
It might be expected that the strategy based on the hierarchy
principle could be helpful in studying the strong coupling limit also
for different theories
(like for instance the strong coupling limit of the Standard
Model \cite{heavyHiggs}-\cite{heavyHiggs6}).

Finally the generality of the approach based
on the hierarchy principle gives some  hope
that it could provide 
an effective strategy for the study of the matching
between the linear and the nonlinear sigma model
in the large mass expansion
also at higher loop orders by preserving  
the unified treatment 
of both the chiral and  the strong coupling limit.

\section*{Acknowledgments}

We thank Tom Appelquist and Juerg Gasser for useful comments.
 
\appendix

\section{Feynman rules for the linear sigma model}\label{app:B}

The classical action $\G^{(0)}$ in eq.(\ref{eq:az2.1}) reads
in terms of the fields $\sigma, \phi_a$
\begin{equation}\label{eq:az3}
\begin{split}
\G^{(0)} =& \int d^Dx \, \Big ( \frac{1}{2}\,\Big(\partial_{\nu}\sigma\,
\partial^{\nu}\sigma\,-\,m^{2}\,\sigma^{2}\,+\,\partial_{\nu}\phi_a
\partial^{\nu}\phi_a\Big)\,-\,\frac{\lambda m ~ v^{2-D/2}}{\sqrt{2}}\,
\Big(\sigma^{3}\,+\,\sigma\,\phi_a^2\Big)\,\\
&~~~~~ \,-\,\frac{\lambda^2 v^{4-D}}{4}\,\Big(\sigma^{4}
\,+2\,\sigma^{2}\,\phi_a^2 \,+\,(\phi_a^2)^{2}
\Big) \\
&~~~~~~ 
+ J_{\mu a} ( -(\sigma + v_D) \,\partial^{\mu}\phi_{a}+\partial^{\mu}
\sigma\,\phi_{a}-\epsilon_{abc}\,\partial^{\mu}\phi_{b} \,
\phi_{c} ) \\
&~~~~~~ + \frac{J_{\mu a}^2}{8} ( \sigma^2 + 2 v_D \sigma +
\phi_a^2 ) + \frac{v_D^2}{8} J_{\mu a}^2  
\Big ) \, .\\[5pt]
\end{split}\end{equation}
In the above equation  we have introduced  the 
mass parameter\, $m~=~\sqrt{2}\,\lambda v$\,.
$\G^{(0)}$ is invariant under the local transformations
\begin{eqnarray}
\delta \phi_a = -\frac{1}{2} (v_D + \sigma) \delta \alpha_a + 
\frac{1}{2} \epsilon_{abc} \phi_b \delta \alpha_c \, , ~~~~
\delta \sigma = \frac{1}{2} \phi_a \delta \alpha_a \, ,
\label{transf.1} 
\end{eqnarray}
while the source $J^\mu_a$ transforms as a background connection
\begin{eqnarray}
\delta J^\mu_a = \partial^\mu \delta \alpha_a + \epsilon_{abc}
J^\mu_b \delta \alpha_c \, .
\label{transf.1.1}
\end{eqnarray}

The tree level Feynman rules 
can be directly read off from eq.(\ref{eq:az3}).
\begin{figure}[p]
\begin{center}
\includegraphics[width=0.4\textwidth]{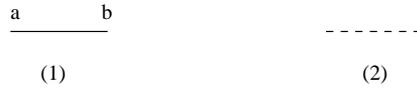}
\end{center}
\caption{Free propagators}
\label{pro}
\end{figure}
\begin{figure}[p]
\begin{center}
\includegraphics[width=0.4\textwidth]{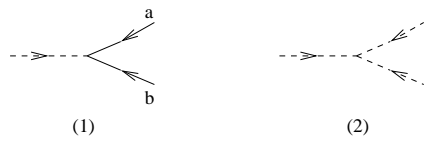}
\end{center}
\caption{Trilinear vertices}
\label{tri}
\end{figure}
\begin{figure}[p]
\begin{center}
\includegraphics[width=0.6\textwidth]{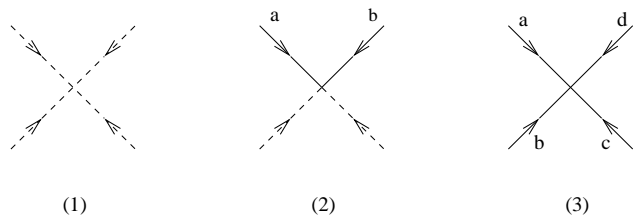}
\end{center}
\caption{Quadrilinear vertices}
\label{qua}
\end{figure}
\begin{figure}[p]
\begin{center}
\includegraphics[width=0.5\textwidth]{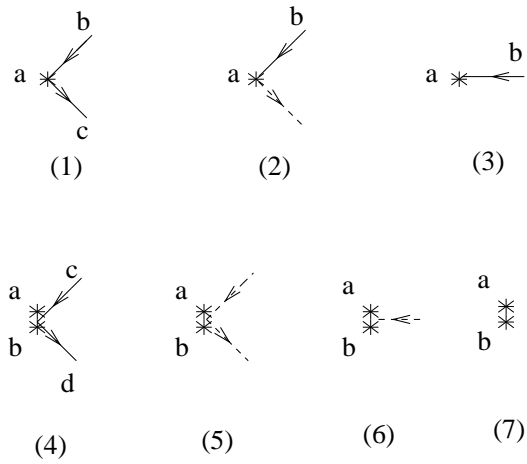}
\end{center}
\caption{Composite operators vertices}
\label{corr}
\end{figure}
\begin{enumerate}
\item Propagators\\
\begin{itemize}
\item Pion propagator\,\, $i\,\Gamma^{(0)-1}_{\phi_{a}\phi_{b}}=\frac{i}
{p^{2}}\,\delta_{ab}$\,\,  (Fig. \ref{pro}.1)\\
\item Sigma propagator\,\, $i\,\Gamma^{(0)-1}_{\sigma\sigma}=\frac{i}
{p^{2}-m^{2}}$\,\, (Fig. \ref{pro}.2)\\
\end{itemize}
\item Trilinear couplings\\
\begin{itemize}
\item $i\,\Gamma^{(0)}_{\phi_{a}\phi_{b}\sigma}=-\frac{2\,i\,
\lambda m ~v^{2-D/2}}{\sqrt{2}}\,\delta_{ab}$\,\, (Fig \ref{tri}.1)\\
\item $i\,\Gamma^{(0)}_{\sigma\sigma\sigma}=-\frac{6\,i\,
\lambda m ~ v^{2-D/2}}{\sqrt{2}}$\,\, (Fig. \ref{tri}.2)\\
\end{itemize}
\item Quadrilinear couplings\\
\begin{itemize}
\item $i\,\Gamma^{(0)}_{\sigma\sigma\sigma\sigma}
=-6\,i\,\lambda^2 v^{4-D}$\,\, (Fig. \ref{qua}.1)\\
\item $i\,\Gamma^{(0)}_{\phi_{a}\phi_{b}\sigma\sigma}=
-2\,i\,\lambda^2 v^{4-D}\,
\delta_{ab}$\,\, (Fig. \ref{qua}.2)\\
\item $i\,\Gamma^{(0)}_{\phi_{a}\phi_{b}\phi_{c}\phi_{d}}
=-2\,i\,\lambda^{2} v^{4-D}\,s_{abcd}$\,\, (Fig. \ref{qua}.3)\\[5pt] 
where\,\, $s_{abcd} =\delta_{ab}\,\delta_{cd}\,+\,
\delta_{ac}\,\delta_{bd}\,+\,\delta_{ad}\,\delta_{bc}$.\\
\end{itemize}
\item Composite operators\\
\begin{itemize}
\item $i\,\Gamma^{(0)}_{\phi_{a}\phi_{b}J_{\mu}^{c}}=
\frac{1}{2}\,\epsilon_{abc}\,\big(p_{1}+p_{2}\big)^{\mu}$\,\, 
(Fig. \ref{corr}.1)\\
\item $i\,\Gamma^{(0)}_{\phi_{a}\sigma J_{\mu b}}=\frac{1}{2}\,\delta_{ab}\,
\big(p_{1}+p_{2}\big)^{\mu}$\,\, (Fig. \ref{corr}.2)\\
\item$i\,\Gamma^{(0)}_{\phi_{a}J_{\mu b}}=\frac{1}{2}\,\delta_{ab}\,v_{D}\,
p^{\mu}\,$\,\, (Fig. \ref{corr}.3)\\
\item $i\,\Gamma^{(0)}_{\phi_{a}\phi_{b} J_{\mu c} J_{\nu d}}=
\frac{i}{2}\,\delta_{ab}\,\delta_{cd}\,g^{\mu\nu}$\,\, (Fig. \ref{corr}.4)\\
\item $i\,\Gamma^{(0)}_{\sigma\sigma J_{\mu a} J_{\nu b}}=
\frac{i}{2}\delta_{ab}\,g^{\mu\nu}$\,\, (Fig. \ref{corr}.5)\\
\item $i\,\Gamma^{(0)}_{\sigma J_{\mu a}J_{\nu b}}=
\frac{i}{2}v_{D}\,\delta_{ab}\,g^{\mu\nu}$\,\, (Fig. \ref{corr}.6)\\
\item $i\,\Gamma^{(0)}_{J_{\mu a}J_{\nu b}}=
\frac{i}{4}v^{2}_{D}\,\delta_{ab}\,g^{\mu\nu}$\,\, (Fig. \ref{corr}.7)\\
\end{itemize}
\end{enumerate}

\section{Useful formulas}
\label{app:C}

In this Appendix we list some useful formulas.\\

\subsection{Massive tadpoles}

We set
\begin{equation}\label{eq:T1}
T_{n}=\int \frac{d^{D}q}{(2\pi)^{D}}\,\frac{1}{\big(q^{2}-m^{2}\big)^{n}}
=\frac{(-)^{n}\,i}{(4\pi)^{D/2}}\,\frac{\Gamma\big(n-D/2\big)}{
\Gamma(n)}\,(m^{2})^{D/2-n}\, .
\end{equation}
The following recursive relation is verified for $T_n$:
\begin{equation}\label{eq:T2}
T_{n}=-\Big(1-\frac{D}{2n-2}\Big)\,m^{-2}\,T_{n-1}\, .
\end{equation}
By properly normalizing $T_n$ one finds for $D \rightarrow 4$
\begin{equation}\label{eq:T4}
v^{2-D} T_1 
=\frac{i}{(4\pi)^{2}}\,\frac{m^{2}}{v^{2}}\,\Big( \frac{2}{4-D} + 1-\gamma_{E}+
\ln(4\pi)-\ln\Big(\frac{m^{2}}{v^{2}}\Big)\Big) \, , 
\end{equation} 
\begin{equation}\label{eq:T5}
v^{4-D} T_{2}=\frac{i}{(4\pi)^{2}}\,\Big( \frac{2}{4-D} -\gamma_{E}+
\ln(4\pi)-\ln\Big(\frac{m^{2}}{v^{2}}\Big)\Big)\, .
\end{equation} 
For\, $n>2$\, we have no more poles in \, $D=4$\,.
The straightforward  
 limit gives
\begin{equation}\label{eq:T6}
T_{n}= \frac{(-)^{n}\,i}{(4\pi)^{2}}\,\frac{1}{(n-1)(n-2)}\,
\frac{1}{m^{2(n-2)}}\, .
\end{equation}

\subsection{Partial fractions identities}

\begin{equation}\label{eq:T7}
I(r,l)=\int \frac{d^{D}q}{(2\pi)^{D}}\,\frac{1}{(q^{2})^{r}\,
(q^{2}-m^{2})^{l}}=m^{-2}\big(I(r-1,l)-I(r,l-1)\big)\, .
\end{equation}
Note that\, $I(0,l)=T_{l}$\, and that in dimensional regularization\, $I(r,0)$
\, vanishes.\, By proceeding recursively one finds\\
\begin{equation}\label{eq:T8}
I(r,l)=\sum_{j=0}^{l-1} \Bigg[(-)^{j}\,\binom{r+j-1}{j}\,m^{-2(r+j)}\,T_{l-j}
\Bigg]\, .
\end{equation}

\subsection{Self-energy integrals}

We collect here the relevant self-energy integrals.
\begin{eqnarray}\label{eq:mi1}
B_0(p^2;0,0) &=& 
v^{4-D}
\int \frac{d^{D}q}{(2\pi)^{D}}\,\frac{1}{q^{2}\,(p+q)^{2}} \nonumber \\
& = & \frac{i}{(4\pi)^{2}}\Big(2-\gamma_{E}+\ln(4\pi)-
\ln\Big(-\frac{p^{2}}{v^{2}}\Big)\Big)\, , 
\end{eqnarray}
\begin{eqnarray}\label{eq:ff2}
B_0(p^2;m,0) & = & v^{4-D}
\int \frac{d^{D}q}{(2\pi)^{D}}\,\frac{1}{(p+q)^{2}(q^{2}-m^{2})}
\nonumber \\
& = & 
\frac{i}{(4\pi)^{2}}\,\Big(1-\gamma_{E}+\ln(4\pi)-\ln\Big(\frac{m^{2}}{v^{2}}
\Big)+f_{1}\Big(\frac{p^{2}}{m^{2}}\Big)\Big)\, , 
\end{eqnarray} 
\begin{eqnarray}
\label{eq:ff4}
B_0(p^2;m,m) & = & v^{4-D} 
\int \frac{d^{D}q}{(2\pi)^{D}}\,\frac{1}{(q^{2}-m^{2})\,[(q+p)^{2}-m^{2}]}
\nonumber \\
& = & 
\frac{i}{(4\pi)^{2}}\Big(-\gamma_{E}+\ln(4\pi)-\ln\Big(\frac{m^{2}}{v^{2}}
\Big)+f_{2}\Big(\frac{p^{2}}{m^{2}}\Big)\Big)\, , 
\end{eqnarray} 
where the functions $f_{1}$ and $f_2$ are given by
\begin{equation}\label{eq:ff3}
f_{1}\Big(\frac{p^{2}}{m^{2}}\Big)=-\int_{0}^{1} dx\,\ln\Big(1-
\frac{p^{2}}{m^{2}}(1-x)\Big)=\frac{p^{2}}{2m^{2}}\Big(1+
\frac{p^{2}}{3m^{2}}+\frac{(p^{2})^{2}}{6m^{4}}\Big)\,+\,O(m^{-8})
\end{equation}
and
\begin{equation}\label{eq:ff6}
f_{2}\Big(\frac{p^{2}}{m^{2}}\Big)=-\int_{0}^{1} dx\,\ln\Big(1-
\frac{p^{2}}{m^{2}}x(1-x)\Big)=\frac{p^{2}}{6m^{2}}\Big(1+
\frac{p^{2}}{10m^{2}}+\frac{\big(p^{2}\big)^{2}}{70m^{4}}\Big)
\,+\,O(m^{-8}) \, .
\end{equation}

\section{One loop four pion function at order $1/m^2$}
\label{app:D}

In this Appendix we verify that the one loop four pion function
$\tilde \G^{(1)}_{\phi_a \phi_b \phi_c \phi_d}$
vanishes at order $1/m^2$. For that purpose we need
to evaluate graphs which are 1-PI  w.r.t. $\phi$ 
and reducible w.r.t. the $\sigma$ lines. This amounts to perform the 
Legendre transform of the 1-PI vertex functional $\G^{(1)}$ w.r.t. $\sigma$.

\begin{figure}[h]
\begin{center}
\includegraphics[width=0.7\textwidth]{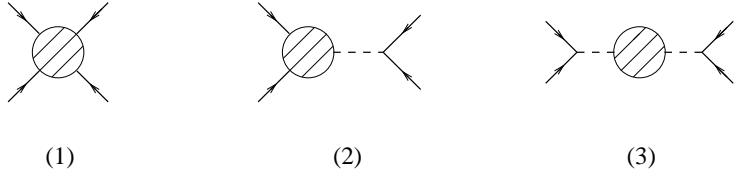}
\end{center}
\caption{Four pion radiative corrections}
\label{quattrog}
\end{figure}

The 1-PI amplitudes of the linear sigma model which enter
in this computation
are $\G^{(1)}_{\phi_a \phi_b \phi_c \phi_d}$,
$\G^{(1)}_{\phi_i \phi_j \sigma}$ and $\G^{(1)}_{\sigma \sigma}$.
They respectively generate the graphs 
 of type 1, 2 and 3 in Fig.~\ref{quattrog}.

In the previous Figure we have not included any graph with the insertion of  
the massive tadpole 
because these graphs are canceled by the counter-term
\,$\delta_{t}$ in eq.(\ref{ren.t}).\\ 

We decompose $\tilde \G^{(1)}_{\phi_a \phi_b \phi_c \phi_d}$
according to the group tensors as follows:
\begin{eqnarray}
\tilde \G^{(1)}_{\phi_a \phi_b \phi_c \phi_d} =
\delta_{ab} \delta_{cd} A+ \delta_{ac} \delta_{bd} B 
+ \delta_{ad} \delta_{bc} C + O(m^{-4}) \, .
\label{appD:decomp}
\end{eqnarray}
We evaluate here the coefficient $A$
as a function of the external momenta and
the parameters of the theory ($B$ and $C$ are obtained
by permutations). Fig.~\ref{quattrog} suggests to further decompose
$A$ according to
\begin{eqnarray}
A = \sum_{i=1}^3 A_i
\label{appD:decomp.1}
\end{eqnarray}
where $A_i$ denotes the contribution to $A$ from the graphs of
type $i$ in Fig.~\ref{quattrog}.

\medskip

\begin{figure}[h]
\begin{center}
\includegraphics[width=0.6\textwidth]{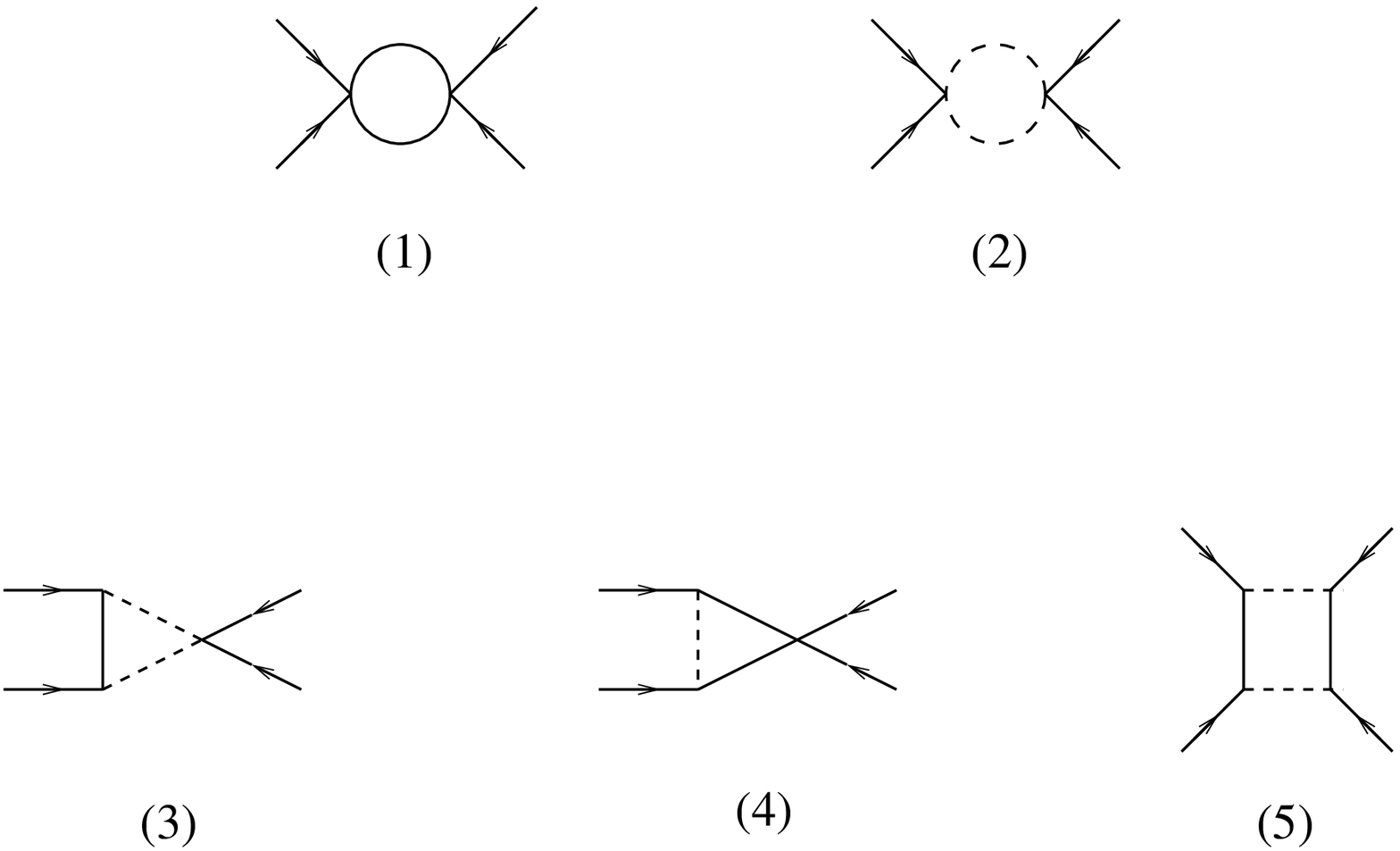}
\end{center}
\caption{1PI}
\label{quattro1}
\end{figure}

We begin by evaluating $A_1$. The relevant 1PI graphs 
are shown in  Fig.~\ref{quattro1}. 
The first two graphs of Fig. \ref{quattro1} can be computed
exactly and yield after Minimal Subtraction\\
\begin{eqnarray}\label{eq:fo1}
A_{1,1} & = & 
\frac{22\lambda^{4}}{(4\pi)^{2}}\Big(2-\gamma_{E}+\ln(4\pi)
-\ln\Big(\frac{m^{2}}{v^{2}}\Big) \Big) \nonumber \\
&& 
\!\!\!\!\!\!\!\!\!
- \frac{14\lambda^{4}}{(4\pi)^{2}}\ln\Big(-\frac{s}{m^{2}}\Big)
- \frac{4\lambda^{4}}{(4\pi)^{2}}\ln\Big(-\frac{t}{m^{2}}\Big)
- \frac{4\lambda^{4}}{(4\pi)^{2}}\ln\Big(-\frac{u}{m^{2}}\Big) 
\end{eqnarray}
(where\, $s=\big(p_{a}+p_{b}\big)^{2}$\, $t=\big(p_{a}+p_{c}\big)^{2}$\,
 $u=\big(p_{a}+p_{d}\big)^{2}$\, are the usual Mandelstam variables)
and
\begin{equation}\label{eq:fo2}
A_{1,2} = 
\frac{2\lambda^{4}}{(4\pi)^{2}}\Big(-\gamma_{E}+\ln(4\pi)-\ln\Big(
\frac{m^{2}}{v^{2}}\Big)+\frac{s}{6m^{2}}\Big) \, . 
\end{equation}
The remaining graphs of Fig.~\ref{quattro1} are UV convergent.
Their direct evaluation for general values of the kinematical invariants is 
not straightforward. However since we are interested in a precise 
kinematical regime  (i.e. the one in which the physical mass parameter\, $m$\, 
is greater than all the others kinematical invariants built with external 
momenta) we can expand the amplitudes in the large $m$ limit
 by making use of a subgraphs asymptotic expansion technique
due to Smirnov \cite{smirnov}-\cite{smirnov2}\,.\\

Here follows a short outline of this technique.\, Given a UV convergent 
 Feynman graph\, ${\cal G}$\, the expansion procedure consists of two steps:
\begin{itemize}
\item Take the whole graph and Taylor-expand it w.r.t. the external 
 momenta around\, $p_{ext}=0$\, . This may introduce spurious IR divergences 
that have to be dimensionally regularized.
\item Take the subgraphs that contain all heavy internal lines 
(i.e. the massive sigma propagators) and are 
1PI w.r.t. light lines (they can be disconnected) and Taylor-expand 
them w.r.t. the external momenta and the momenta flowing in the heavy 
 lines around\, $p_{ext}=0$\, and\, $q_{heavy}=0$\,.\,This may introduce 
 spurious UV divergences that have to be dimensionally regularized.
\end{itemize}    
It can be proven that for any order in the expansion and for any graph the 
spurious divergences exactly cancel each other.
The remaining terms provide the large mass expansion of the
amplitude ${\cal G}$.

By applying this technique we find that 
graph (3) of Fig. \ref{quattro1} yields
\begin{equation}\label{eq:fo7}
A_{1,3} = -\frac{4\lambda^{4}}{(4\pi)^{2}} 
\Big ( 2 + \frac{5s + 3t + 3u}{12m^{2}} \Big )\,.\\
\end{equation}
The contribution from graph (4) in Fig.~\ref{quattro1} is
\begin{equation}\label{eq:fo11}\begin{split}
\!\!\!\!\!\!
A_{1,4} = &\frac{4\lambda^{4}}{(4\pi)^{2}}
\Bigg[2\Big(\ln\Big(-\frac{s}{m^{2}}\Big)
 + \ln\Big(-\frac{t}{m^{2}}\Big) + \ln\Big(-\frac{u}{m^{2}}\Big) - 3 \Big)
 + \frac{5}{4}\frac{s + t + u}{m^{2}}\\
& +\frac{-s+t+u}{2m^{2}}\ln\Big(-\frac{s}{m^{2}}\Big)+
\frac{s-t+u}{2m^{2}}\ln\Big(-\frac{t}{m^{2}}\Big)+
\frac{s+t-u}{2m^{2}}\ln\Big(-\frac{u}{m^{2}}\Big)\Bigg]\, .
\end{split}\end{equation}
Finally we discuss the Feynman integral associated to graph (5) of 
Fig \ref{quattro1}:
\begin{eqnarray}\label{eq:fo14}
\!\!\!\!\!\!
A_{1,5} & = &
-\frac{4\lambda^{4}}{(4\pi)^{2}}\,\Bigg[\ln\Big(-\frac{t}{m^{2}}\Big)\,+\,
\ln\Big(-\frac{u}{m^{2}}\Big)\,+\,
\frac{5s}{3m^{2}}\,+\,\frac{t+u}{m^{2}}\,+\nonumber \\
&& 
\!\!\!\!\!\!\!\!\!\!\!\!\!\!\!\!\!
+\frac{s-t+u}{2m^{2}}\Big(\ln\Big(-\frac{t}{m^{2}}\Big)-1\Big)+
\frac{s+t-u}{2m^{2}}\Big(\ln\Big(-\frac{u}{m^{2}}\Big)-1\Big)\Bigg]\,.
\end{eqnarray}
By collecting together the results obtained 
in eqs.(\ref{eq:fo1}), (\ref{eq:fo2}) and 
 (\ref{eq:fo7})-(\ref{eq:fo14}) we find the contribution of the 1PI 
 graphs:
\begin{eqnarray}\label{eq:fo15}
A_1 & = & \frac{24\lambda^{4}}{(4\pi)^{2}}
\Big(\frac{1}{2} - \gamma_{E}+\ln(4\pi)-\ln\Big(\frac{m^{2}}{v^{2}}\Big) 
- \frac{1}{4} \ln\Big(-\frac{s}{m^{2}}\Big) \Big)
 + \frac{\lambda^{4}}{(4\pi)^{2}}\frac{s}{m^{2}}\nonumber \\
&& +\frac{4\lambda^{4}}{(4\pi)^{2}}\frac{-s+t+u}{2m^{2}}
\ln\Big(-\frac{s}{m^{2}}\Big)\,.
\end{eqnarray}

\begin{figure}[h]
\begin{center}
\includegraphics[width=0.6\textwidth]{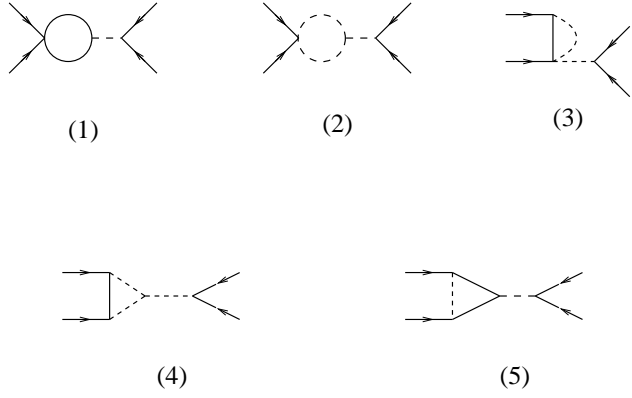}
\end{center}
\caption{One sigma line}
\label{quattro2}
\end{figure}

Now we consider the graphs of type 2 (displayed in Fig.~\ref{quattro2}).
Graph (1) gives
\begin{eqnarray}
\!\!\!\!\!\!\!\!\!\!\!
A_{2,1} & = & -\frac{20 \lambda^4}{(4\pi)^2} 
\Big ( 1 + \frac{s}{m^2} \Big ) 
\Big ( 2 + \gamma_E - \ln(4\pi) - \ln \Big ( \frac{m^{2}}{v^{2}} \Big )
               - \ln \Big (-\frac{s}{m^2} \Big ) \Big ) \, .
\label{appD:a21}
\end{eqnarray}
Graph (2) gives
\begin{eqnarray}
\!\!\!\!\!\!
A_{2,2} & = & -\frac{12 \lambda^4}{(4\pi)^2} 
\Big ( 1 + \frac{s}{m^2} \Big ) 
\Big ( \gamma_E - \ln(4\pi) - \ln\Big(\frac{m^{2}}{v^{2}} \Big ) \Big)
              - \frac{2 \lambda^4}{(4\pi)^2} \frac{s}{m^2} \, .
\label{appD:a22}
\end{eqnarray}
The contribution from graph (3) in Fig.~\ref{quattro2} is 
\begin{eqnarray}
\!\!\!\!\!\!\!\!\!\!\!\!\!
A_{2,3} & = & -\frac{16 \lambda^4}{(4\pi)^2} 
\Big ( 1 + \frac{s}{m^2} \Big ) 
\Big ( \gamma_E - \ln(4\pi) - \ln\Big(\frac{m^{2}}{v^{2}} \Big ) \Big ) 
              - \frac{2 \lambda^4}{(4\pi)^2} \frac{s + t + u}{m^2} \, .
\label{appD:a23}
\end{eqnarray}
Graph (4) gives
\begin{eqnarray}
\!\!\!\!\!\!
A_{2,4} & = & \frac{24 \lambda^4}{(4\pi)^2} \Big ( 1 + \frac{s}{m^2} \Big ) 
             + \frac{\lambda^4}{(4\pi)^2} \frac{5 s + 3 t + 3 u}{m^2} \, .
\label{appD:a24}
\end{eqnarray}
Finally graph (5) gives
\begin{eqnarray}\label{appD:a25}
\!\!\!\!\!\!
A_{2,5} & = &-\frac{8\lambda^{4}}{(4\pi)^{2}}\Big ( 1 + \frac{s}{m^2} \Big ) 
          \Big( \ln\Big (-\frac{s}{m^{2}} \Big ) - 1 \Big) \nonumber \\ 
         && -\frac{2\lambda^{4}}{(4\pi)^{2}} \Big ( \frac{3 s + t + u}{2m^{2}}
 +\frac{-s + t + u}{m^{2}} \ln \Big (-\frac{s}{m^{2}} \Big ) \Big ) \, .
\end{eqnarray}
By collecting together the results obtained 
in eqs.(\ref{appD:a21})-(\ref{appD:a25}) 
we find the contribution of the graphs of type 2:
\begin{eqnarray}\label{eq:fo16}
A_2 & = &  -\frac{48\lambda^{4}}{(4\pi)^{2}} 
\Big( \frac{1}{2}-\gamma_{E}+\ln(4\pi)
-\ln\Big(\frac{m^{2}}{v^2}\Big) - \frac{1}{4} 
\ln\Big(-\frac{s}{m^{2}}\Big) \Big)
\nonumber \\
&& -\frac{48\lambda^{4}}{(4\pi)^{2}} \frac{s}{m^2} 
\Big( \frac{13}{24}-\gamma_{E}+\ln(4\pi)
-\ln\Big(\frac{m^{2}}{v^2}\Big) - \frac{1}{4} 
\ln\Big(-\frac{s}{m^{2}}\Big) \Big) 
 \nonumber \\
&& -\frac{2\lambda^{4}}{(\pi)^{2}} \frac{-s + t + u}{m^{2}} 
\ln \Big (-\frac{s}{m^{2}} \Big ) \, .
\end{eqnarray} 

\begin{figure}[h]
\begin{center}
\includegraphics[width=0.6\textwidth]{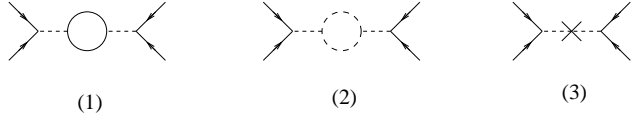}
\end{center}
\caption{Two sigma lines}
\label{quattro3}
\end{figure}

Finally we consider the graphs of type 3 (displayed in Fig.~\ref{quattro3}).
Graph (1) gives
\begin{eqnarray}
\!\!\!\!\!\!\!\!\!\!\!
A_{3,1} & = & \frac{6 \lambda^4}{(4\pi)^2} \Big ( 1 + \frac{2s}{m^2} \Big ) 
               \Big ( 2 + \gamma_E - \ln(4\pi) - 
\ln \Big ( \frac{m^{2}}{v^{2}} \Big )
               - \ln \Big (-\frac{s}{m^2} \Big ) \Big ) \, .
\label{appD:a31}
\end{eqnarray}
The contribution from graph (2) in Fig.~\ref{quattro3} is
\begin{eqnarray}
\!\!\!\!\!\!
A_{3,2} & = & \frac{18 \lambda^4}{(4\pi)^2} 
\Big ( 1 + \frac{2s}{m^2} \Big ) 
\Big ( \gamma_E - \ln(4\pi) - \ln\Big(\frac{m^{2}}{v^{2}} \Big ) \Big)
              + \frac{3 \lambda^4}{(4\pi)^2} \frac{s}{m^2} \, .
\label{appD:a32}
\end{eqnarray}
Finally the contribution from graph (3), controlled
by the wave function counterterm $\delta_Z$, yields
\begin{eqnarray}\label{eq:con4}
&& 2\lambda^2\,\delta_Z\,\frac{s}{m^2}\Big(1-\frac{s}{m^2}\Big)^{-2}\, = 
-\frac{2\lambda^4}{(4\pi)^2}\frac{s}{m^{2}} + O(m^{-4})\, .
\end{eqnarray} 

By collecting together the results obtained 
in eqs.(\ref{appD:a31})-(\ref{eq:con4})
we find the contribution of the graphs of type 3:
\begin{eqnarray}\label{eq:fo17}
A_3 & = &  \frac{24\lambda^{4}}{(4\pi)^{2}} 
\Big( \frac{1}{2}-\gamma_{E}+\ln(4\pi)
-\ln\Big(\frac{m^{2}}{v^{2}}\Big) - \frac{1}{4} 
\ln\Big(-\frac{s}{m^{2}}\Big) \Big)
\nonumber \\
&& \!\!\!\!\!\!
+ \frac{48\lambda^{4}}{(4\pi)^{2}} \frac{s}{m^2} 
\Big( \frac{27}{48}-\gamma_{E}+\ln(4\pi)-\ln\Big(\frac{m^{2}}{v^{2}}\Big) - 
\frac{1}{4} \ln\Big(-\frac{s}{m^{2}}\Big) \Big) \nonumber \\
&& \!\!\!\!\!\! 
-\frac{2\lambda^4}{(4\pi)^2}
\frac{s}{m^{2}} \, .
\end{eqnarray} 
By summing the contributions in eqs.(\ref{eq:fo15}), 
(\ref{eq:fo16}) and (\ref{eq:fo17}) it can be checked 
that the terms of order $1/m^2$ exactly
cancel. Therefore the first non-vanishing term for the four-point
function $\tilde \G^{(1)}_{\phi_a \phi_b \phi_c \phi_d}$ 
is of order $1/m^4$.

Several comments are in order here. First we remark
that the non-local contributions containing logs of the momenta
cancel each other among eqs.(\ref{eq:fo15}), 
(\ref{eq:fo16}) and (\ref{eq:fo17}). On the other hand, if one does not impose
the renormalization condition on the residue of the pion
in eq.(\ref{ren.z}) the four-point
function $\tilde \G^{(1)}_{\phi_a \phi_b \phi_c \phi_d}$ 
would contain a term of order $1/m^2$
\begin{eqnarray}
\frac{2\lambda^4}{(4\pi)^2}
\frac{s}{m^{2}}
\label{appD:cts}
\end{eqnarray}
which diverges in the strong
coupling limit and goes like $1/v^2$ in the chiral limit (in contrast with the 
 corresponding leading terms of the nonlinear model which are of the type 
$1/v^4 \ln v$), thus spoiling in both cases the matching with the 
nonlinear theory.

\section{Superficially divergent ancestor amplitudes}
\label{app:G}

In this Appendix we evaluate 
diagrammatically the contributions
$\Delta \tilde \G^{(1)}_{K_0 K_0}$, 
$\Delta \tilde \G^{(1)}_{JJ}$, $\Delta
\tilde \G^{(1)}_{K_0 JJ}$,$\Delta \tilde \G^{(1)}_{JJJ}$
 and
$\Delta \tilde \G^{(1)}_{JJJJ}$
to the superficially divergent
ancestor amplitudes at the first order in the large mass expansion
from graphs with at least one sigma line.

The subset of the graphs which are 1-PI w.r.t. the sigma lines
allows us to evaluate the contribution to the amplitudes
$\G^{(1)}_{\sigma \sigma}, \G^{(1)}_{\sigma JJ}$,
$\G^{(1)}_{JJ}, \G^{(1)}_{JJJ}$ and $\G^{(1)}_{JJJJ}$
from graphs with massive internal lines
in the strong coupling limit, thus fixing
$\G^{(1)}_H$ in eq.(\ref{appBB:1}).

\begin{itemize}
\item $\Delta \tilde \G^{(1)}_{K_0 K_0}$

The relevant graphs come from  the radiative corrections 
to the sigma propagator. 
At one loop level there are two of them (Fig. \ref{dues}).\\
\begin{figure}[h]
\begin{center}
\includegraphics[width=0.5\textwidth]{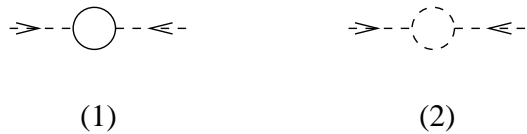}
\end{center}
\caption{Sigma propagator radiative corrections}
\label{dues}
\end{figure}
Graph (1) coincides exactly with the corresponding graph of the 
nonlinear sigma model and thus need not be 
computed.
Graph (2) gives
\begin{equation}\label{eq:two4}
\Delta \tilde\Gamma_{K_0 K_0}^{(1)}(p) = - \frac{9}{2(4\pi)^{2}v^2}
\Big(-\gamma_{E} + \\
\ln(4\pi) - \ln\Big(\frac{m^2}{v^2}\Big)\Big)\, .
\end{equation} 
The contribution of graph (2) to the two-point sigma function
in $\G^{(1)}_H$ is
\begin{eqnarray}
\G^{(1)}_{H,\sigma\sigma} = - \frac{18 \lambda^4 v^2}{(4\pi)^2} 
\ln \frac{m^2}{v^2}\, .
\label{appG:e1}
\end{eqnarray}
\item $\Delta \tilde \G^{(1)}_{JJ}$
 
We now move to the graphs involving two external currents.
\begin{figure}[h!]
\begin{center}
\includegraphics[width=0.7\textwidth]{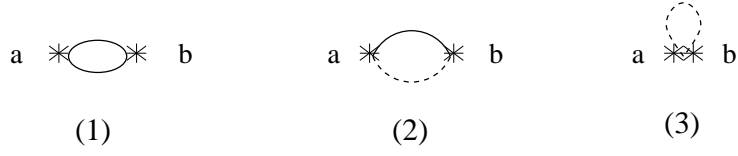}
\caption{Two currents}
\label{corr3}
\end{center}
\end{figure}
Graph (1) in Fig.~\ref{corr3}  coincides exactly with the
corresponding graph of the nonlinear sigma model.
Graph (2) gives
\begin{eqnarray}\label{eq:one10}
\Delta \tilde\Gamma_{J^{\mu}_{a} J^{\nu}_{b},2}^{(1)}(p) &=& 
\Big[\frac{\lambda^{2}v^{2}}{2(4\pi)^{2}}\Big(\frac{3}{2}-\gamma_{E}+
\ln(4\pi)-\ln\Big(\frac{m^2}{v^2}\Big)\Big) 
g_{\mu\nu} \nonumber \\
&& + \frac{1}{12(4\pi)^{2}}\Big(\frac{4}{3} -
\gamma_{E}+\ln(4\pi)-\ln\Big(\frac{m^2}{v^2}\Big)\Big) 
p_\mu p_\nu  \nonumber \\
&& - \frac{1}{12(4\pi)^{2}}\Big(\frac{5}{6}-\gamma_{E} + 
\ln(4\pi)-\ln\Big(\frac{m^2}{v^2}\Big)\Big)
p^{2}g_{\mu\nu}\Big] \delta_{ab} \, . \nonumber \\
\end{eqnarray} 
The contribution of graph (3) is
\begin{equation}\label{eq:one12}
\Delta \tilde\Gamma^{(1)}_{J^{\mu}_{a} J^{\nu}_{b},3}(p) = 
-\frac{\lambda^{2}v^{2}}
{2(4\pi)^{2}}\,\Big(1-\gamma_{E} +
\ln(4\pi)-\ln\Big(\frac{m^2}{v^2}\Big)\Big)\,g_{\mu\nu}\delta_{ab}\, .
\end{equation}
Finally we need to take into account the
 contribution from the
counterterm proportional to $\delta_Z$ in eq.(\ref{ren.2}):
\begin{eqnarray}
\tilde\G^{(1)}_{J^{\mu}_{a} J^{\nu}_{b},ct}(p) = -
\frac{\lambda^{2}v^{2}}{4(4\pi)^{2}}g_{\mu\nu}\delta_{ab}\, .
\label{appG:jjct}
\end{eqnarray}
Thus
\begin{eqnarray}
\Delta \tilde \G^{(1)}_{J^{\mu}_{a} J^{\nu}_{b}}(p)  &=&
\frac{1}{12(4\pi)^{2}} \Big [\Big(\frac{4}{3} -\gamma_{E} + 
\ln(4\pi)-\ln\Big(\frac{m^2}{v^2}\Big)\Big) p_\mu p_\nu  \nonumber \\
&& - \Big(\frac{5}{6}-\gamma_{E}+\ln(4\pi)-\ln\Big(\frac{m^2}{v^2}\Big)\Big)
p^{2}g_{\mu\nu}\Big] \delta_{ab} \, .
\label{appG:jj}
\end{eqnarray}

Moreover
\begin{eqnarray}
\G^{(1)}_{H,J^{\mu}_{a} J^{\nu}_{b}}
& = & - \frac{1}{12(4\pi)^{2}} \ln\Big(\frac{m^2}{v^2}\Big)
(p_\mu p_\nu - g_{\mu\nu} p^2) \, .
\label{appG:jj1}
\end{eqnarray}

\item $\Delta \tilde \G^{(1)}_{K_0 JJ}$

Now we consider the graphs with two external currents 
and one $K_0$.
First we compute the set of graphs in Fig.~\ref{corr8}.
\begin{figure}[h!]
\begin{center}
\includegraphics[width=0.7\textwidth]{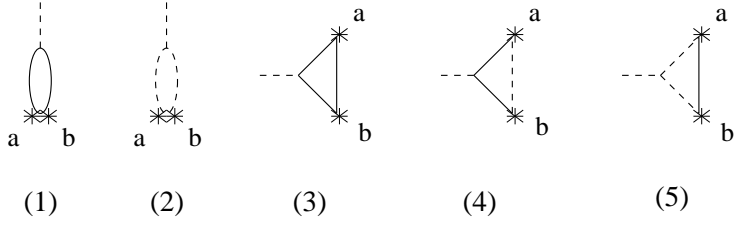}
\end{center}
\caption{Two currents and one $K_0$ (1-PI)}
\label{corr8}
\end{figure}  
Graph (1) gives
\begin{equation}\label{eq:one11}
\Delta \tilde\Gamma_{K_0 J^{\mu}_{a} J^{\nu}_{b},1}^{(1)}(p,p_1,p_2)= -\frac{3}
{4(4\pi)^{2}v}\Big(2-\gamma_E +\ln(4\pi)-
\ln\Big(-\frac{p^2}{v^2}\Big)\Big) g_{\mu\nu} \delta_{ab}\, .
\end{equation} 
Graph (2) gives
\begin{equation}\label{eq:one13}
\Delta \tilde\Gamma_{K_0 J^{\mu}_{a} J^{\nu}_{b},2}^{(1)}(p,p_1,p_2)= -\frac{3}
{4(4\pi)^{2}v}\Big(-\gamma_E + \ln(4\pi) -
\ln\Big(\frac{m^2}{v^2}\Big)\Big) g_{\mu\nu} \delta_{ab}\, .
\end{equation}
Graph (3) has not to be computed
since it coincides with the corresponding one
of the nonlinear sigma model.
Graph (4) gives
\begin{equation}\label{eq:one14}
\Delta \tilde\Gamma_{K_0 J^{\mu}_{a} J^{\nu}_{b},4}^{(1)}(p,p_1,p_2)= \frac{1}
{4(4\pi)^{2}v}\Big(\frac{3}{2}-\gamma_E + \ln(4\pi) -
\ln\Big(\frac{m^2}{v^2}\Big)\Big) g_{\mu\nu} \delta_{ab}\, .
\end{equation}
Finally graph (5) gives
\begin{equation}\label{eq:one15}
\Delta \tilde\Gamma_{K_0 J^{\mu}_{a} J^{\nu}_{b},5}^{(1)}(p,p_1,p_2)= \frac{3}
{4(4\pi)^{2}v}\Big(\frac{1}{2}-\gamma_E + \ln(4\pi) -
\ln\Big(\frac{m^2}{v^2}\Big)\Big) g_{\mu\nu} \delta_{ab}\, .
\end{equation}
There are also graphs which are not 1-PI, depicted in
Fig.~\ref{corr6}.
\begin{figure}[h!]
\begin{center}
\includegraphics[width=0.5\textwidth]{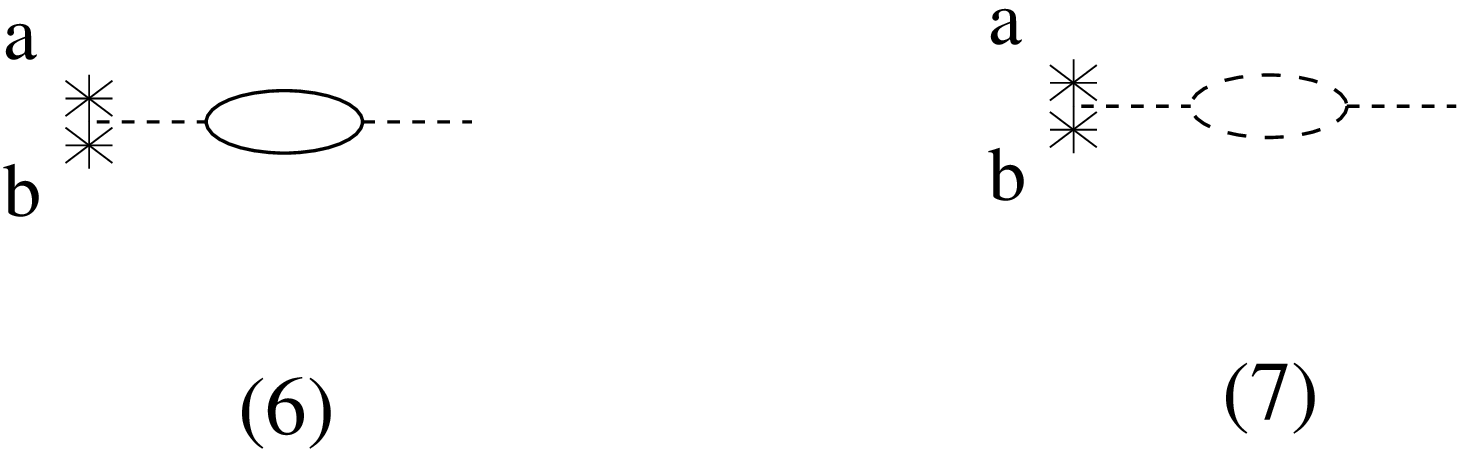}
\end{center}
\caption{Two currents and one $K_0$ with one sigma line}
\label{corr6}
\end{figure}  
Graph (6) gives
\begin{equation}\label{eq:one16}
\Delta \tilde\Gamma_{K_0 J^{\mu}_{a} J^{\nu}_{b},6}^{(1)}(p,p_1,p_2)= \frac{3}
{4(4\pi)^{2}v}\Big(2-\gamma_E +\ln(4\pi)-
\ln\Big(-\frac{p^2}{v^2}\Big)\Big) g_{\mu\nu} \delta_{ab}\, .
\end{equation} 
Graph (7) 
\begin{equation}\label{eq:one17}
\Delta \tilde\Gamma_{K_0 J^{\mu}_{a} J^{\nu}_{b},7}^{(1)}(p,p_1,p_2)= \frac{9}
{4(4\pi)^{2}v}\Big(-\gamma_E + \ln(4\pi) -
\ln\Big(\frac{m^2}{v^2}\Big)\Big) g_{\mu\nu} \delta_{ab}\, .
\end{equation}
Hence we obtain
\begin{equation}
\Delta \tilde \G^{(1)}_{K_0 J^{\mu}_{a} J^{\nu}_{b}}  =  
\frac{1}{4 (4\pi)^2 v} \Big [ 3 - 10 \gamma_E + 10 \ln (4 \pi)
- 10 \ln \Big ( \frac{m^2}{v^2} \Big ) \Big ] g_{\mu\nu} \delta_{ab} \, .
\label{appD:kjj1}
\end{equation}
The contribution to $\G^{(1)}_H$ is obtained by
taking into account only the 1-PI contributions in
Fig.~\ref{corr8}. Thus
\begin{eqnarray}
\G^{(1)}_{H,\sigma J_\mu^a J_\nu^b} =
- \frac{\lambda^2 v}{2 (4 \pi)^2} \ln \Big ( \frac{m^2}{v^2} \Big )
g_{\mu\nu} \delta_{ab} \, .
\label{appD:kij2}
\end{eqnarray}

\item $\Delta \tilde \G^{(1)}_{JJJ}$

Now we consider the graphs with three external currents (Fig. \ref{corr4}).
\begin{figure}[h!]
\begin{center}
\includegraphics[width=0.4\textwidth]{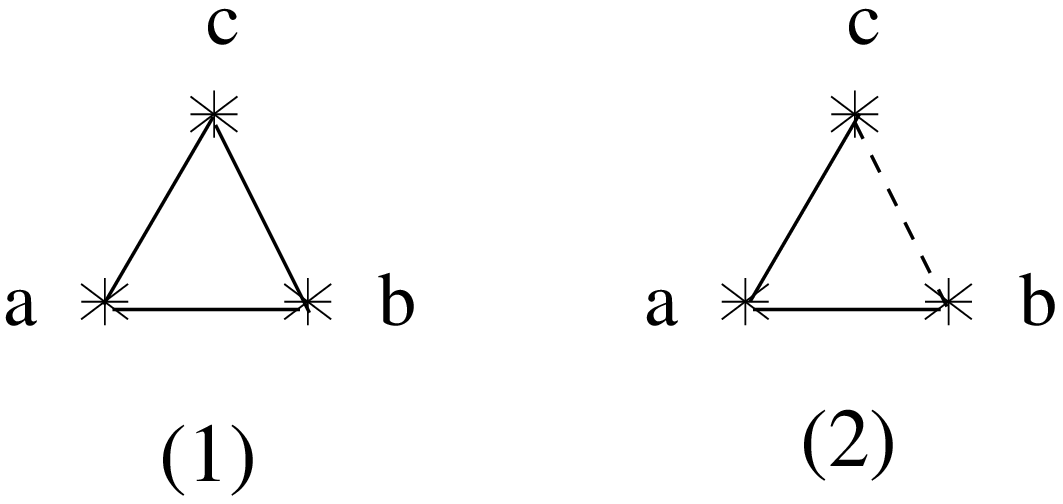}
\end{center}
\caption{Three currents}
\label{corr4}
\end{figure}  
Graph 1 coincides exactly with 
the corresponding Feynman graph of the nonlinear sigma model.

The contribution of the second graph is:
\begin{eqnarray}\label{eq:three1}
&& \!\!\!\!\!\!\!\!\!\!\!\!\!\!\!\!\!\!\!\!\!\!\!
\Delta \tilde\Gamma^{(1)}_{J^{\mu}_{a} J^{\nu}_{b} J^{\rho}_{c}}(p_1,p_2,p_3)=
-\frac{i \epsilon_{abc}}{8(4\pi)^2} 
\Big(\frac{3}{2}-\gamma_{E}+
\ln(4\pi)-\ln\Big(\frac{m^2}{v^2}\Big)\Big)
\nonumber \\
&& ~~~~~ \big[g_{\mu\nu}(p_1-p_2)_{\rho}
-g_{\mu\rho}(p_1-p_3)_{\nu}+g_{\nu\rho}(p_2-p_3)_{\mu}\big]\, ,
\end{eqnarray} 
and therefore
\begin{eqnarray}
&& \!\!\!\!\!\!\!\!\!\!\!\!\!\!\!\!\!\!\!\!\!\!\!
\Gamma^{(1)}_{H,J^{\mu}_{a} J^{\nu}_{b} J^{\rho}_{c}}(p_1,p_2,p_3)=
\frac{i \epsilon_{abc}}{8(4\pi)^2} 
\ln\Big(\frac{m^2}{v^2}\Big)
\nonumber \\
&& ~~~~~ \big[g_{\mu\nu}(p_1-p_2)_{\rho}
-g_{\mu\rho}(p_1-p_3)_{\nu}+g_{\nu\rho}(p_2-p_3)_{\mu}\big] \, .
\label{JJJh}
\end{eqnarray}

\item $\Delta \tilde \G^{(1)}_{JJJJ}$
  
Finally we consider the graphs with four external currents 
(Fig. \ref{corr5}).\\
\begin{figure}[h!]
\begin{center}
\includegraphics[width=0.7\textwidth]{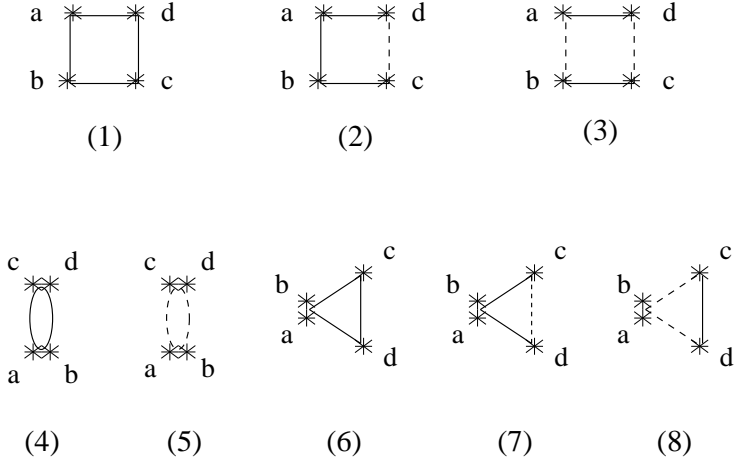}
\end{center}
\caption{Four currents 1-PI}
\label{corr5}
\end{figure} 
Graph (1) coincides exactly with 
the corresponding graph of the nonlinear sigma model.
Graph (2) vanishes at leading order.
Graph (3) gives
\begin{eqnarray}\label{eq:four9}
\Delta \tilde\Gamma^{(1)}_{J^{\mu_1}_{a_1} J^{\mu_2}_{a_2} J^{\mu_3}_{a_3}
J^{\mu_4}_{a_4},3}(p_1,p_2,p_3,p_4)&=&\frac{s_{a_1 a_2 a_3 a_4}}{12(4\pi)^2} 
g_{(\mu_1 \mu_2}g_{\mu_3\mu_4)}\nonumber\\
&& \!\!\!\!\!\!\!\!\!\!\!\!\!\!\!\!\!\!\!\!\!\!
\Big(\frac{5}{6}-\gamma_{E}+\ln(4\pi)-\ln\Big(\frac{m^2}{v^2}\Big)\Big)
\end{eqnarray}
where 
\begin{eqnarray}
g_{(\mu_1 \mu_2}g_{\mu_3\mu_4)} = g_{\mu_1 \mu_2}g_{\mu_3\mu_4}
+ g_{\mu_1 \mu_3}g_{\mu_2\mu_4} + g_{\mu_1 \mu_4}g_{\mu_2\mu_3} \, .
\label{symm}
\end{eqnarray}
Graph (4) gives
\begin{eqnarray}\label{eq:four10}
&& 
\!\!\!\!\!\!\!\!\!\!\!\!\!\!
\Delta \tilde\Gamma^{(1)}_{J^{\mu_1}_{a_1} J^{\mu_2}_{a_2} J^{\mu_3}_{a_3}
J^{\mu_4}_{a_4},4}(p_1,p_2,p_3,p_4)  = - \frac{3}{8(4\pi)^2}\nonumber \\
&& \Big[g_{\mu_1 \mu_2}g_{\mu_3\mu_4}
\delta_{a_1 a_2}\delta_{a_3 a_4}
\Big ( -2+\gamma_E -\ln(4\pi) + 
\ln\Big(-\frac{s}{v^2}\Big) \Big )\nonumber \\
&& + g_{\mu_1 \mu_3}g_{\mu_2\mu_4}\delta_{a_1 a_3}\delta_{a_2 a_4}
\Big ( -2+\gamma_E -\ln(4\pi) +
\ln\Big(-\frac{t}{v^2}\Big) \Big ) 
\nonumber\\
&& + g_{\mu_1 \mu_4}g_{\mu_2\mu_3}
\delta_{a_1 a_4}\delta_{a_2 a_3}
\Big ( -2+\gamma_E -\ln(4\pi) +
\ln\Big(-\frac{u}{v^2}\Big)\Big] \, .
\end{eqnarray}
Graph (5) gives
\begin{equation}\label{eq:four11}\begin{split}
\!\!\!\!\!\!
\Delta \tilde\Gamma^{(1)}_{J^{\mu_1}_{a_1} J^{\mu_2}_{a_2} J^{\mu_3}_{a_3}
J^{\mu_4}_{a_4},5}(p_1,p_2,p_3,p_4)=\frac{1}{8(4\pi)^2} & \big(g_{\mu_1 \mu_2}
g_{\mu_3\mu_4}\delta_{a_1 a_2}\delta_{a_3 a_4}+ 
{\tiny
\begin{matrix} (1 \leftrightarrow 3) \\ (2 \leftrightarrow 4) \end{matrix}
}
+{\tiny
\begin{matrix} (1 \leftrightarrow 4) \\ (2 \leftrightarrow 3) \end{matrix}
}
)\\
& \Big(-\gamma_{E}+\ln(4\pi)-\ln\Big(\frac{m^2}{v^2}\Big)\Big)\, .
\end{split}\end{equation}

At the leading order
graph (6) cancels exactly against graph (11) of Fig. \ref{corr9}.

Graph (7) gives
\begin{eqnarray}\label{eq:four12}
\!\!\!\!\!\!\!
\Delta \tilde\Gamma^{(1)}_{J^{\mu_1}_{a_1} J^{\mu_2}_{a_2} J^{\mu_3}_{a_3}
J^{\mu_4}_{a_4},7}(p_1,p_2,p_3,p_4)
& = & -\frac{1}{4(4\pi)^2}
\big(g_{\mu_1 \mu_2}
g_{\mu_3\mu_4}\delta_{a_1 a_2}\delta_{a_3 a_4}+ 
{\tiny
\begin{matrix} (1 \leftrightarrow 3) \\ (2 \leftrightarrow 4) \end{matrix}
}
+{\tiny
\begin{matrix} (1 \leftrightarrow 4) \\ (2 \leftrightarrow 3) \end{matrix}
}
)
\nonumber \\
&& \Big(\frac{3}{2}-\gamma_{E}+\ln(4\pi)-\ln\Big(\frac{m^2}{v^2}\Big)\Big)\, .
\end{eqnarray}
Graph (8) gives
\begin{eqnarray}\label{eq:four13}
\!\!\!\!\!\!\!
\Delta \tilde\Gamma^{(1)}_{J^{\mu_1}_{a_1} J^{\mu_2}_{a_2} J^{\mu_3}_{a_3}
J^{\mu_4}_{a_4},8}(p_1,p_2,p_3,p_4) & = &-\frac{1}{4(4\pi)^2}
\big(g_{\mu_1 \mu_2}
g_{\mu_3\mu_4}\delta_{a_1 a_2}\delta_{a_3 a_4}+ 
{\tiny
\begin{matrix} (1 \leftrightarrow 3) \\ (2 \leftrightarrow 4) \end{matrix}
}
+{\tiny
\begin{matrix} (1 \leftrightarrow 4) \\ (2 \leftrightarrow 3) \end{matrix}
}
)
\nonumber \\
&& \Big(\frac{1}{2}-\gamma_{E}+\ln(4\pi)-\ln\Big(\frac{m^2}{v^2}\Big)\Big)\, .
\end{eqnarray}

\medskip

Now we move to the connected graphs with one sigma line attached
to a 1-PI one loop amplitude (see Fig.~\ref{corr9}).
\begin{figure}[h!]
\begin{center}
\includegraphics[width=0.8\textwidth]{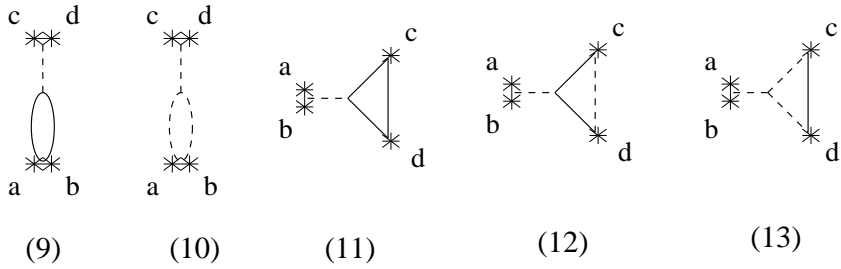}
\end{center}
\caption{Four currents with one sigma line}
\label{corr9}
\end{figure} 
Graph (9) gives
\begin{eqnarray}\label{eq:four14}
&& \!\!\!\!\!\!\!\!\!\!\!\!\!\!
\Delta \tilde\Gamma^{(1)}_{J^{\mu_1}_{a_1} J^{\mu_2}_{a_2} J^{\mu_3}_{a_3}
J^{\mu_4}_{a_4},9}(p_1,p_2,p_3,p_4)
=  \frac{3}{4(4\pi)^2}\nonumber \\
&& \Big[
g_{\mu_1 \mu_2}g_{\mu_3\mu_4}
\delta_{a_1 a_2}\delta_{a_3 a_4}
\Big ( -2+\gamma_E -\ln(4\pi) + 
\ln\Big(-\frac{s}{v^2}\Big) \Big )\nonumber \\
&& + g_{\mu_1 \mu_3}g_{\mu_2\mu_4}\delta_{a_1 a_3}\delta_{a_2 a_4}
\Big ( -2+\gamma_E -\ln(4\pi) +
\ln\Big(-\frac{t}{v^2}\Big) \Big ) 
\nonumber\\
&& + g_{\mu_1 \mu_4}g_{\mu_2\mu_3}
\delta_{a_1 a_4}\delta_{a_2 a_3}
\Big ( -2+\gamma_E -\ln(4\pi) +
\ln\Big(-\frac{u}{v^2}\Big)\Big] \, .
\end{eqnarray}
Graph (10) gives
\begin{equation}\label{eq:four15}\begin{split}
\!\!\!\!\!\!
\Delta \tilde\Gamma^{(1)}_{J^{\mu_1}_{a_1} J^{\mu_2}_{a_2} J^{\mu_3}_{a_3}
J^{\mu_4}_{a_4},10}(p_1,p_2,p_3,p_4) = 
-\frac{3}{4(4\pi)^2} & \big(g_{\mu_1 \mu_2}
g_{\mu_3\mu_4}\delta_{a_1 a_2}\delta_{a_3 a_4}+ 
{\tiny
\begin{matrix} (1 \leftrightarrow 3) \\ (2 \leftrightarrow 4) \end{matrix}
}
+{\tiny
\begin{matrix} (1 \leftrightarrow 4) \\ (2 \leftrightarrow 3) \end{matrix}
}
)\\
& \Big(-\gamma_{E}+\ln(4\pi)-\ln\Big(\frac{m^2}{v^2}\Big)\Big)\, .
\end{split}\end{equation}
Graph (12) gives
\begin{equation}\label{eq:four16}\begin{split}
\!\!\!\!\!\!
\Delta \tilde\Gamma^{(1)}_{J^{\mu_1}_{a_1} J^{\mu_2}_{a_2} J^{\mu_3}_{a_3}
J^{\mu_4}_{a_4},12}(p_1,p_2,p_3,p_4)=\frac{1}{4(4\pi)^2} & \big(g_{\mu_1 \mu_2}
g_{\mu_3\mu_4}\delta_{a_1 a_2}\delta_{a_3 a_4}+ 
{\tiny
\begin{matrix} (1 \leftrightarrow 3) \\ (2 \leftrightarrow 4) \end{matrix}
}
+{\tiny
\begin{matrix} (1 \leftrightarrow 4) \\ (2 \leftrightarrow 3) \end{matrix}
}
)\\
& \Big(\frac{3}{2} -\gamma_{E}+\ln(4\pi)-\ln\Big(\frac{m^2}{v^2}\Big)\Big)\, .
\end{split}\end{equation}
Graph (13) gives
\begin{equation}\label{eq:four17}\begin{split}
\!\!\!\!\!\!
\Delta \tilde\Gamma^{(1)}_{J^{\mu_1}_{a_1} J^{\mu_2}_{a_2} J^{\mu_3}_{a_3}
J^{\mu_4}_{a_4},13}(p_1,p_2,p_3,p_4)=\frac{3}{4(4\pi)^2} & \big(g_{\mu_1 \mu_2}
g_{\mu_3\mu_4}\delta_{a_1 a_2}\delta_{a_3 a_4}+ 
{\tiny
\begin{matrix} (1 \leftrightarrow 3) \\ (2 \leftrightarrow 4) \end{matrix}
}
+{\tiny
\begin{matrix} (1 \leftrightarrow 4) \\ (2 \leftrightarrow 3) \end{matrix}
}
)\\
& \Big(\frac{1}{2}-\gamma_{E}+\ln(4\pi)-\ln\Big(\frac{m^2}{v^2}\Big)\Big)\, .
\end{split}\end{equation}

\medskip

Finally we consider the connected graphs with two sigma lines
attached to a 1-PI amplitude (see Fig.~\ref{corr7}).
\begin{figure}[h!]
\begin{center}
\includegraphics[width=0.6\textwidth]{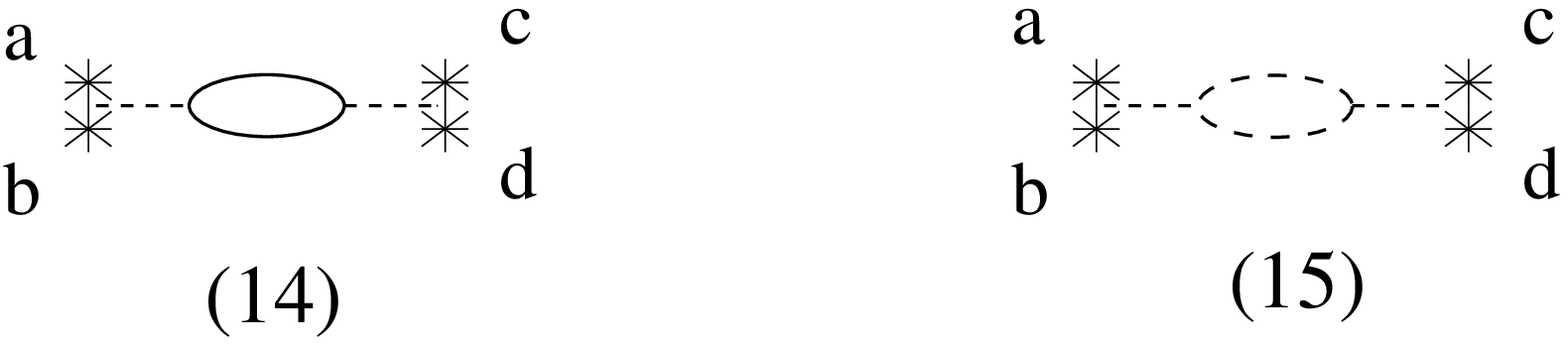}
\end{center}
\caption{Four currents with two sigma lines}
\label{corr7}
\end{figure}
Graph (14) gives\\
\begin{eqnarray}\label{eq:four18}
&& 
\!\!\!\!\!\!\!\!\!\!\!\!\!\!\!\!
\Delta \tilde\Gamma^{(1)}_{J^{\mu_1}_{a_1} J^{\mu_2}_{a_2} J^{\mu_3}_{a_3}
J^{\mu_4}_{a_4},14}(p_1,p_2,p_3,p_4)  = - \frac{3}{8(4\pi)^2}\nonumber \\
&& \!\! \Big[g_{\mu_1 \mu_2}g_{\mu_3\mu_4}
\delta_{a_1 a_2}\delta_{a_3 a_4}
\Big ( -2+\gamma_E -\ln(4\pi) + 
\ln\Big(-\frac{s}{v^2}\Big) \Big )\nonumber \\
&& \!\! + g_{\mu_1 \mu_3}g_{\mu_2\mu_4}\delta_{a_1 a_3}\delta_{a_2 a_4}
\Big ( -2+\gamma_E -\ln(4\pi) +
\ln\Big(-\frac{t}{v^2}\Big) \Big ) 
\nonumber\\
&& \!\! + g_{\mu_1 \mu_4}g_{\mu_2\mu_3}
\delta_{a_1 a_4}\delta_{a_2 a_3}
\Big ( -2+\gamma_E -\ln(4\pi) +
\ln\Big(-\frac{u}{v^2}\Big)\Big] \, .
\end{eqnarray}

We remark that 
the amplitudes in eqs.(\ref{eq:four10}), (\ref{eq:four14}) 
and (\ref{eq:four18})
cancel among each other. This is a consequence of the
general cancellation mechanism implemented by eq.(\ref{appBB:7}).
Notice that 
in the case of eqs.(\ref{eq:four10}), (\ref{eq:four14}) and (\ref{eq:four18})
this cancellation guarantees the absence
of terms containing logs of the momenta in the corrections
to the ancestor amplitudes from graphs with at least one massive line
(at the leading order in the large $m$ expansion).

Finally graph (15) gives
\begin{equation}\label{eq:four19}\begin{split}
\!\!\!\!\!\!
\Delta \tilde\Gamma^{(1)}_{J^{\mu_1}_{a_1} J^{\mu_2}_{a_2} J^{\mu_3}_{a_3}
J^{\mu_4}_{a_4},15}(p_1,p_2,p_3,p_4)=\frac{9}{8(4\pi)^2} & \big(g_{\mu_1 \mu_2}
g_{\mu_3\mu_4}\delta_{a_1 a_2}\delta_{a_3 a_4}+ 
{\tiny
\begin{matrix} (1 \leftrightarrow 3) \\ (2 \leftrightarrow 4) \end{matrix}
}
+{\tiny
\begin{matrix} (1 \leftrightarrow 4) \\ (2 \leftrightarrow 3) \end{matrix}
}
)\\
& \Big(-\gamma_{E}+\ln(4\pi)-\ln\Big(\frac{m^2}{v^2}\Big)\Big)\, .
\end{split}\end{equation}
Hence we are left with 
\begin{eqnarray}
\Delta \tilde \G^{(1)}_{J^{\mu_1}_{a_1} J^{\mu_2}_{a_2} J^{\mu_3}_{a_3}
J^{\mu_4}_{a_4}} & = & 
\frac{s_{a_1 a_2 a_3 a_4}}{12(4\pi)^2} 
g_{(\mu_1 \mu_2}g_{\mu_3\mu_4)}
\Big(\frac{5}{6}-\gamma_{E}+\ln(4\pi)-\ln\Big(\frac{m^2}{v^2}\Big)\Big)
\nonumber \\
&& + 
\frac{1}{(4\pi)^2}
\big(g_{\mu_1 \mu_2}
g_{\mu_3\mu_4}\delta_{a_1 a_2}\delta_{a_3 a_4}+ 
{\tiny
\begin{matrix} (1 \leftrightarrow 3) \\ (2 \leftrightarrow 4) \end{matrix}
}
+{\tiny
\begin{matrix} (1 \leftrightarrow 4) \\ (2 \leftrightarrow 3) \end{matrix}
}
)\nonumber \\
&& ~~~~ \Big [ \frac{1}{4} - \gamma_E + \ln (4\pi) - \ln
\Big( \frac{m^2}{v^2} \Big ) \Big ] \, .
\label{4j.fin}
\end{eqnarray}
By considering only the contributions from the 1-PI
graphs in Fig.~\ref{corr5}
we can evaluate $\G^{(1)}_{H,J^{\mu_1}_{a_1} J^{\mu_2}_{a_2} J^{\mu_3}_{a_3}
J^{\mu_4}_{a_4}}$:
\begin{eqnarray}
\G^{(1)}_{H,J^{\mu_1}_{a_1} J^{\mu_2}_{a_2} J^{\mu_3}_{a_3}
J^{\mu_4}_{a_4}} & = & -
\frac{s_{a_1 a_2 a_3 a_4}}{12(4\pi)^2} 
g_{(\mu_1 \mu_2}g_{\mu_3\mu_4)}
\ln\Big(\frac{m^2}{v^2}\Big)
\nonumber \\
&& 
\!\!\!\!\!\!\!\!\!\!\!\!
+ \frac{3}{8(4\pi)^2}
\big(g_{\mu_1 \mu_2}
g_{\mu_3\mu_4}\delta_{a_1 a_2}\delta_{a_3 a_4}+ 
{\tiny
\begin{matrix} (1 \leftrightarrow 3) \\ (2 \leftrightarrow 4) \end{matrix}
}
+{\tiny
\begin{matrix} (1 \leftrightarrow 4) \\ (2 \leftrightarrow 3) \end{matrix}
}
)\ln\Big(\frac{m^2}{v^2}\Big) 
\, .\nonumber \\
\label{4jH.fin}
\end{eqnarray}
\end{itemize}

\section{Nonlinear sigma model}
\label{app:E}

The $D$-dimensional classical action of the nonlinear sigma model in the
flat connection formalism \cite{Ferrari:2005ii} is
\begin{eqnarray}
&& \G^{(0)}_{NL} =  \frac{v_D^2}{8} \int d^Dx \, 
\Big ( F^\mu_a - J^\mu_a \Big )^2 + \int d^Dx \, K_0 \phi_0 \, . 
\label{appE:1}
\end{eqnarray}
$J_\mu^a$ is the background connection and $K_0$ the source
of the constraint $\phi_0$ of the nonlinear sigma model
\begin{eqnarray}
\phi_0^2 + \phi_j^2 = v_D^2 \, .
\label{appE:2}
\end{eqnarray}
$\G^{(0)}_{NL}$ obeys the following $D$-dimensional
local functional equation
\begin{eqnarray}
&& 
+ \frac{1}{2} \frac{\delta \G^{(0)}_{NL}}{\delta K_0} 
\frac{\delta \G^{(0)}_{NL}}{\delta \phi_a}
+\frac{1}{2} \epsilon_{abc}\phi_c \frac{\delta \G^{(0)}_{NL}}{\delta \phi_b} 
- \frac{1}{2} \phi_a K_0 \nonumber \\
&& - \partial^\mu \frac{\delta \G^{(0)}_{NL}}{\delta J^\mu_a} 
   -  \epsilon_{abc} J^\mu_b 
\frac{\delta \G^{(0)}_{NL}}{\delta J^\mu_c} = 0 \, .
\label{appE:2bis}
\end{eqnarray}

We set $S_0 = \left . \G^{(0)}_{NL} \right |_{K_0=0}$.
In terms of the background connection $J_\mu^a$ and of the
flat connection
\begin{eqnarray}
F^\mu_a = \frac{2}{v_D^{2}}\big(\phi_0 \partial^\mu \phi_a - 
\partial^\mu \phi_0 \phi_a+ \epsilon_{abc} \partial^\mu \phi_b \phi_c\big) \, 
\label{appE:3}
\end{eqnarray}
the invariant solutions of the linearized functional equation 
which enter at the one loop level read \cite{Ferrari:2005va}
\begin{eqnarray}
&& {\cal I}_1 = \int d^Dx \, 
\Big [ D_\mu ( F -J )_\nu \Big ]_a \Big [ D^\mu ( F -J )^\nu \Big ]_a  \, , 
\nonumber \\
&& {\cal I}_2 = \int d^Dx \, 
\Big [ D_\mu ( F -J )^\mu \Big ]_a \Big [ D_\nu ( F -J )^\nu \Big ]_a  \, , 
\nonumber \\
&& {\cal I}_3 = \int d^Dx \, 
\epsilon_{abc} \Big [ D_\mu ( F -J )_\nu 
\Big ]_a \Big ( F^\mu_b -J^\mu_b \Big ) 
\Big ( F^\nu_c -J^\nu_c \Big ) \, ,  \nonumber \\
&& {\cal I}_4 = \int d^Dx \, 
\Big ( \frac{v_D^2 K_0}{\phi_0} - 
\phi_a \frac{\delta S_0}{\delta \phi_a} \Big )^2 \, , \nonumber \\
&& {\cal I}_5 = \int d^Dx \, 
\Big ( \frac{v_D^2 K_0}{\phi_0} - 
\phi_a \frac{\delta S_0}{\delta \phi_a} \Big ) 
\Big ( F^\mu_b -J^\mu_b \Big )^2 \, , 
\nonumber \\
&& {\cal I}_6 = \int d^Dx \, \Big ( F^\mu_a -J^\mu_a\Big  )^2
 \Big ( F^\nu_b -J^\nu_b \Big )^2 \, , \nonumber \\
&& {\cal I}_7 = \int d^Dx \, \Big ( F^\mu_a -J^\mu_a\Big  )
   \Big ( F^\nu_a -J^\nu_a\Big  ) \nonumber \\
&& ~~~~~~~~~~~~~~~~
   \Big ( F_{b\mu} -J_{b\mu} \Big  )
   \Big ( F_{b\nu} -J_{b\nu} \Big  ) \, ,
\label{appE:4}
\end{eqnarray}
where $D_\mu$ denotes the covariant derivative w.r.t $F_{\mu a}$:
\begin{eqnarray}
D_{\mu,ab} = \partial_\mu \delta_{ab} + \epsilon_{acb} F_{\mu c} \, .
\label{appE:5}
\end{eqnarray}


\begin{thebibliography}{30}
%
\bibitem{Ferrari:2005ii}
  R.~Ferrari,
  JHEP {\bf 0508} (2005) 048
  [arXiv:hep-th/0504023].
%

\bibitem{Ferrari:2005va}
  R.~Ferrari and A.~Quadri,
  Int.\ J.\ Theor.\ Phys.\  {\bf 45} (2006) 2497
  [arXiv:hep-th/0506220].

\bibitem{Ferrari:2005fc}
  R.~Ferrari and A.~Quadri,
  JHEP {\bf 0601} (2006) 003
  [arXiv:hep-th/0511032].
%
\bibitem{Bessis:1972sn}
  D.~Bessis and J.~Zinn-Justin,
  Phys.\ Rev.\ D {\bf 5} (1972) 1313.
%
\bibitem{Appelquist:1980ae}
  T.~Appelquist and C.~W.~Bernard,
  Phys.\ Rev.\ D {\bf 23} (1981) 425.
%
\bibitem{Akhoury:1982hv}
  R.~Akhoury and Y.~P.~Yao,
  Phys.\ Rev.\ D {\bf 25} (1982) 3361.
%
\bibitem{Sonoda:1996pd}
  H.~Sonoda,
  Nucl.\ Phys.\ B {\bf 490} (1997) 457
  [arXiv:hep-th/9609132].
%
\bibitem{smirnov}
  V.~A.~Smirnov,
  Commun.\ Math.\ Phys.\  {\bf 134} (1990) 109.

\bibitem{smirnov1}
  V.~A.~Smirnov,
  Mod.\ Phys.\ Lett.\ A {\bf 10} (1995) 1485
  [arXiv:hep-th/9412063].

\bibitem{smirnov2}
  V.~A.~Smirnov,
  ``Applied asymptotic expansions in momenta and masses,''
  Springer Tracts Mod.\ Phys.\  {\bf 177} (2002).
%
\bibitem{Picariello:2000xc}
  M.~Picariello and A.~Quadri,
  Phys.\ Lett.\ B {\bf 497} (2001) 91
  [arXiv:hep-th/0001174].
%
\bibitem{Quadri:2005pv}
  A.~Quadri,
  JHEP {\bf 0506} (2005) 068
  [arXiv:hep-th/0504076].
%
\bibitem{Quadri:2003pq}
  A.~Quadri,
  J.\ Phys.\ G {\bf 30} (2004) 677
  [arXiv:hep-th/0309133].
%
\bibitem{Quadri:2003ui}
  A.~Quadri,
  JHEP {\bf 0304} (2003) 017
  [arXiv:hep-th/0301211].
%
\bibitem{CHPT}
  J.~Gasser and H.~Leutwyler,
  Annals Phys.\  {\bf 158}, 142 (1984).

\bibitem{CHPT1}
  A.~Nyffeler and A.~Schenk,
  Annals Phys.\  {\bf 241}, 301 (1995)
  [arXiv:hep-ph/9409436].
  
\bibitem{heavyHiggs}
  T.~Appelquist and C.~W.~Bernard,
  Phys.\ Rev.\ D {\bf 22} (1980) 200.

\bibitem{heavyHiggs1}
  A.~C.~Longhitano,
  Nucl.\ Phys.\ B {\bf 188} (1981) 118.

\bibitem{heavyHiggs2}
  M.~J.~Herrero and E.~Ruiz Morales,
  Nucl.\ Phys.\ B {\bf 418} (1994) 431
  [arXiv:hep-ph/9308276].

\bibitem{heavyHiggs3}
  M.~J.~Herrero and E.~Ruiz Morales,
  Nucl.\ Phys.\ B {\bf 437} (1995) 319
  [arXiv:hep-ph/9411207].

\bibitem{heavyHiggs4}
  S.~Dittmaier and C.~Grosse-Knetter,
  Nucl.\ Phys.\ B {\bf 459}, 497 (1996)
  [arXiv:hep-ph/9505266].

\bibitem{heavyHiggs5}
  S.~Dittmaier and C.~Grosse-Knetter,
  Phys.\ Rev.\ D {\bf 52}, 7276 (1995)
  [arXiv:hep-ph/9501285].

\bibitem{heavyHiggs6}
  R.~Ferrari, M.~Picariello and A.~Quadri,
  Phys.\ Lett.\ B {\bf 611} (2005) 215
  [arXiv:hep-th/0409194].

%


\end{thebibliography}
\end{document}